\def\Roman#1{\uppercase\expandafter{\romannumeral#1}}
\documentclass[a4paper,12pt]{article}
\usepackage{cite}
\usepackage{amssymb,amsfonts}
\usepackage[mathscr]{eucal}
\usepackage{mathrsfs}
\usepackage[utf8x]{inputenc}

\title{The Lagrange-Poincaré equations for interacting Yang-Mills and scalar fields}

\author{S. N. Storchak\footnote{E-mail adress: storchak@ihep.ru}\\
\small{  NRC ``Kurchatov Institute'' -- IHEP,}\\
\small{Protvino, Moscow Region,  142281,   Russia}}

\begin{document}

\maketitle

\begin{abstract}
A special case of the Lagrange-Poincaré equations for the gauge field interacting with a scalar field is obtained. For description of the dynamics on the configuration space, the adapted coordinates are used. After neglecting the group variables the obtained equations describe the evolution on the gauge orbit space of the principal fiber bundle which is   related to the  system under the consideration.

\end{abstract}

\section{Introduction}
The behavior of systems with symmetry is determined by internal dynamics, which is often hidden, which presents significant difficulties in the case of the usual description of evolution.
In the theory of reduction for mechanical systems with symmetry, this problem is solved using the Lagrange-Poincaré equations. Due to symmetry, the configuration space of mechanical systems  can be regarded as the total space of the principal fiber bundle associated with the system. 
The Lagrange-Poincare
equations are given by two equations: the ``horizontal'' equation which belongs to the kernel of the 1-form connection (naturally emergent in such systems) and the ``vertical'' equation related to  the motion along the orbit of the principal fiber bundle.

In case of the projection onto the base manifold (the orbit space of the principal fiber bundle), the horizontal equation describes the internal dynamics of the system. This dynamics is determined by the mechanical system that arises from the original system as a result of the reduction. In mechanics, the interrelation between the original system and the reduced one is well studied due to  Marsden-Weinstein reduction theory.\cite{AbrMarsd,Marsden}  
But internal dynamics is also the main object of research in gauge theories — infinite-dimensional dynamic systems that are invariant with respect to the action of a group of gauge transformations. Here, the true configuration space (the configuration space of physically observable variables) is the orbit space of the action of the gauge group. The main problem for these systems is that it is not possible to describe the local dynamics on the gauge orbit space in terms of the gauge-invariant variables. It is currently unknown how to do this in satisfactory way.
 
The generally accepted method of describing local dynamics in orbit space is to use a special coordinate system in the principal fiber bundle. The coordinates of such a  system are known by the name of the adapted coordinates\cite{Narasimhan, Daniel-Viallet, Mitter-Viallet,Babelon-Viallet} and are defined using local sections of the bundle. The sections are given  by local surfaces (submanifolds) in the  space of gauge fields.
The local surfaces themselves are determined by equations that cannot be explicitly resolved, so parametric representations of the surfaces cannot be obtained. As a result, when introducing coordinates into the principal bundle, we are forced to deal with constrained  variables (or dependent variables) as  coordinates in this approach. In spite of this, the approach is widely used, for example, when quantizing  gauge fields by the path integral method.\cite{Creutz, Gawedzki, Huffel-Kelnhofer} Studies of the classical evolution of gauge fields with the use of adapted coordinates for local descriptions of the dynamics have practically not been conducted.

In this paper our goal is to obtain the Lagrange-Poincare equations for the gauge system formed from the Yang-Mills field interacting with the scalar field.  We are based on our works\cite{Stor_lagr_poinc_1,Stor_lagr_poinc_2} where we have considerd the mechanical system of two particals given on the product manifold consisting of the Riemannian manifold and the manifold represented by the vector space. It was assumed that the system under consideration is invariant with respect to the group action. The resulting reduced mechanical system was given on the corresponding associated bundle which serves as the base space of the principal bundle related to the system. The geometry of this special mechanical system is analogous to the gauge system we consider in the present article. So it can be regarded as the model system for our problem.
 
The paper will be organized as follows.  Section 2 is an introduction to our paper, where we recall  our previous  work from arXiv, where the mechanical system of two interacting particles was investigated. In Section 3 we  explain how the adapted coordinates can be determined for the gauge interacting sistem formed from the Yang-Mills field and a scalar field. These coordinates correspond to the coordinates in the mechanical system. This provides the basis for using the Lagrange-Poincaré equations obtained earlier for the mechanical system, in deriving analogous equations for the gauge system under the study. In Section 3, we derive such equations for the gauge system using functional expressions for the terms of the Lagrange – Poincaré equations  obtained earlier for the mechanical system. Details of derivations of the Lagrange-Poincar\'{e} equations are considered in Appendix.

\section{Mechanical system of two interacting particles}
In our previous works \cite{Stor_lagr_poinc_1, Stor_lagr_poinc_2}, we considered a special finite-dimensional mechanical system with the following Lagrangian:
\begin{equation}  
\mathcal L=\frac12 G_{AB}(Q)\,{\dot Q}^A{\dot Q}^B +\frac12 G_{mn}\,{\dot f}^m{\dot f}^n-V(Q,f).
\label{lagrang_1}
\end{equation}
The configuration space of this system is  the product manifold $\mathcal P \times V$. It was assumed that $\mathcal P$ is a smooth finite-dimensional  Riemannian manifold (without the boundary) and $V$ is a finite-dimensional vector space. So,  
$(Q^A,f^n)$, $A=1,\dots , N_P$ and $n=1,\dots ,N_V$, are the coordinates of a point $(p,v)\in \mathcal P \times V$ in some local chart. 
Also, it was assumed that  a smooth isometric free and proper action of the compact group Lie $\mathcal G$ on $\mathcal P \times V$ was given. We dealt with  the right action on $\mathcal P \times V$: $(p,v)g=(pg,g^{-1}v)$. In coordinates, this action is written as follows:
\[
 {\tilde Q}^A=F^A(Q,g),\;\;\;\;{\tilde f}^n=\bar D^n_m(g)f^m.
\]
Here $\bar D^n_m(g)\equiv D^n_m(g^{-1})$,
and by $D^n_m(g)$ we denote the matrix of  the finite-dimensional representation of the group $\mathcal G$
acting on the vector space $V$.

For our metric
\begin{equation}
 ds^2=G_{AB}(Q)dQ^AdQ^B+G_{mn}df^mdf^n,
\label{metr_orig}
\end{equation}
the Killing vector fields 
$$
K_{\alpha}(Q,f)= K^B_{\alpha}(Q)\frac{\partial}{\partial Q^B}+K^p_{\alpha}\frac{\partial}{\partial { f}^p}$$
have the following components: 
$K^B_{\alpha}(Q)=\frac{\partial {\tilde Q}^B}{\partial a^{\alpha}}\Big|_{a=e}$ and
$K^p_{\alpha}(f)=({\bar J}_{\alpha})^p_m f^m$.  (The generators ${\bar J}_{\alpha}$ of the representation ${\bar D}^n_m(a)$ are such that
$[{\bar J}_{\alpha},{\bar J}_{\beta}]={\bar c}^{\gamma}_{\alpha \beta}{\bar J}_{\gamma}$, where 
${\bar c}^{\gamma}_{\alpha \beta}=-{c}^{\gamma}_{\alpha \beta}$.)

In the following, we will also use the condensed notation for indices: 
$\tilde A\equiv (A,p)$. So, for example,  the components of the Killing vector fields will be written as 
$ K^{\tilde A}_{\mu}=(K^{A}_{\mu},K^{p}_{\mu}).$ 

From the general theory\cite{AbrMarsd} it is known that in our case $\mathcal P \times V$ can be regarded as a total space of the principal fiber bundle
\[
 \pi': \mathcal P \times V\to \mathcal P\times _{\mathcal G}V,
\]
that is, $\pi': (p,v)\to [p,v]$, where  $[p,v]$ is the equivalence class with respect to the relation $(p,v)\sim (pg,g^{-1}v)$.

Due to this fact it is possible to express the coordinates $(Q^A, f^n)$ of the point $(p,v)$ in terms of the  the principal fiber bundle coordinates. The method of performing this for the typical principal bundle $\rm P(M,\mathcal G)$ is well-known \cite{Creutz, Razumov, Storchak_11, Storchak_12,Storchak_2,Storchak_3}. In approach close to ours was considered in \cite{Huffel-Kelnhofer} for the abelian gauge theory.  
It consists of using the local sections $\tilde \sigma_i$ of our  bundle, $\pi' \cdot\tilde \sigma_i = \rm{id}$. But to define $\tilde \sigma_i$, it is necessary to use the sections $ \sigma_i$ of the principal fiber bundle $\rm P(\mathcal M,\mathcal G)$:
\[
 \tilde \sigma_i([p,v])=(\sigma_i(x),a(p) v)=(\tilde p,\tilde v)\in \mathcal P\times V,
\]
where $a(p)$ is the group element defined by $p=\sigma_i(x)a(p)$.

The adapted coordinates on  $\rm P(\mathcal M,\mathcal G)$ are defined by means of the choice of the special local sections $\sigma_i$. The sections are determined by the local submanifold $\Sigma_i$ of $\mathcal P$, given by the equation $\{\chi ^{\alpha}(Q)=0,\alpha =1,\dots,N_{\mathcal G}\}$.
The  coordinates  of the points on the local submanifold $\Sigma_i$ 
will be denoted by $Q^{\ast}{}^A$, they are such that $\{\chi^{\alpha}(Q^{\ast})=0\}$. That is, 
$Q^{\ast}{}^A$ are dependent coordinates.
In other words, the special section $\sigma_i$ is defined as  the map $\sigma_i:U_i\to \Sigma_i$: $\pi_{\Sigma_i}\cdot\sigma_i={\rm id}_{U_i}$.

We note that there exists  a local isomorphism between trivial principal bundle $\Sigma_{i}\times \mathcal G\to\Sigma_{i}$ and 
$\rm P(\mathcal M,\mathcal G)$:\cite{Mitter-Viallet,Babelon-Viallet,Huffel-Kelnhofer}  
\[
 \varphi_{i}:\,\,\Sigma_{i}\times \mathcal G\to \pi ^{-1}(U_{i}),
\]
which allows us  to introduce a local coordinates on $\rm P(\mathcal M,\mathcal G)$.
In coordinates we have:
\[
 \varphi_{i}:(Q^{\ast}{}^B,a^{\alpha})\to Q^A=F^A(Q^{\ast}{}^B, a^{\alpha}),
\]
where $Q^{\ast}{}^B$ are the coordinates of a  point given  on the local surface $ \Sigma_{i}$ and 
$a^{\alpha}$ -- the coordinates of an arbitrary group element $a$. This element carries the point, taken on $ \Sigma_{i}$, to  the point $p\in \mathcal P$ which has  the coordinates $Q^A$.

An inverse map $\varphi_{i}^{-1}$,
\[
 \varphi_{i}^{-1}:\,\,\pi ^{-1}(U_{i})\to\Sigma_{i}\times \mathcal G,
\]
has the following coordinate representation:
\[
 \varphi_{i}^{-1}: Q^A\to (Q^{\ast}{}^B(Q),a^{\alpha}(Q)).
\]
Here the group coordinates $a^{\alpha}(Q)$ of a point $p$ are the coordinates of the group element  
which connects, by means of its  action on $p$, the surface $\Sigma _{i}$ and the point $p\in \mathcal P$. These group coordinates are given by the solutions of the following equation:
\begin{equation}
 \chi^{\beta}(F^A(Q, a^{-1}(Q)))=0.
\label{a_chi}
\end{equation} 
The coodinates $Q^{\ast}{}^B$ are defined by the equation
\begin{equation}
 Q^{\ast}{}^B=F^B(Q, a^{-1}(Q)).
\label{Q_star}
\end{equation}

In the same way as for the principal bundle $\rm P(\mathcal M,\mathcal G)$, there exist a local isomorphisms of the principal fiber bundle ${\rm P}(\mathcal P\times _{\mathcal G}V,\mathcal G)$ and the  trivial principal bundles $\tilde \Sigma_i\times \mathcal G \to \tilde \Sigma_i$, where now the local surfaces $\tilde \Sigma_i$ are the images of the sections $\tilde \sigma_i$.

In this case we have the following coordinate functions  of the charts:
\[
 \tilde \varphi_i^{-1} : \pi^{-1}(\tilde U_i)\to\tilde \Sigma_i \times \mathcal G,\;\;{\rm or}\; {\rm  in}\; {\rm coordinates,} 
\]
\[
 \tilde \varphi_i^{-1} :(Q^A,f^m)\to (Q^{\ast}{}^A(Q),\tilde f^n(Q),a^{\alpha}(Q)\,).
\]
Here  $Q^A$ and $f^m$ are the coordinates of a point $(p,v)\in \mathcal P\times V$,  
$Q^{\ast}{}^A(Q)$ is given by (\ref{Q_star}) and
\[
\tilde f^n(Q) = D^n_m(a(Q))\,f^m,
\]
$a(Q)$ is defined by  (\ref{a_chi}), and  we have used the following property: $\bar D^n_m(a^{-1})\equiv D^n_m(a)$.  The coordinates $Q^{\ast}{}^A$, representing  a point given on a local surface $\Sigma_i$,  satisfy the constraints: $\chi(Q^{\ast})=0$.

The coordinate function $\tilde \varphi_i$ maps $\tilde \Sigma_i\times \mathcal G\to \pi^{-1}(\tilde U_i)$:
\[
 \tilde \varphi_i :(Q^{\ast}{}^B,\tilde f^n,a^{\alpha})\to (F^A(Q^{\ast},a), \bar D^m_n(a)\tilde f^n).
\]
Thus, we have determined the special  local bundle coordinates $(Q^{\ast}{}^A,\tilde f^n, a^{\alpha})$, also called the  adapted coordinates, 
in the principal fiber bundle 
$\pi:\mathcal P\times V\to \mathcal P\times_{\mathcal G} V$.

The replacement of the coordinate basis $\displaystyle(\partial /\partial Q^B,\partial/\partial a^{\alpha})$ for a new basis 
$\displaystyle(\partial/\partial Q{}^{\ast A},\partial/\partial \tilde f^m,\partial/\partial a^{\alpha})$ is performed as follows:
\begin{eqnarray}
\displaystyle 
&&\!\!\!\!\!\!\!\!\frac{\partial}{\partial f^n}=D^m_n(a)\frac{\partial}{\partial {\tilde f}^m},
\nonumber\\
&&\!\!\!\!\!\!\!\!\frac{\partial}{\partial Q^B}=\frac{\partial Q^{\ast}{}^A}{ \partial Q^B}\frac{\partial}{\partial Q^{\ast}{}^A}+\frac{\partial a^{\alpha}}{\partial Q^B}\frac{\partial}{\partial a^{\alpha}}+\frac{\partial {\tilde f}^m}{\partial Q^B}\frac{\partial}{\partial {\tilde f}^m}
\nonumber\\
&&\!\!\!\!\!\!\!\!\!\!\!=\check F^C_B\Biggl(N^A_C(Q^{\ast})\frac{\partial}{\partial Q^{\ast}{}^A}+{\chi}^{\mu}_C({\Phi}^{-1})^{\beta}_{\mu}\bar{v}^{\alpha}_{\beta}(a)\frac{\partial}{\partial a^{\alpha}}-{\chi}^{\mu}_C({\Phi}^{-1})^{\nu}_{\mu}(\bar J_{\nu})^m_p\tilde f^p\frac{\partial}{\partial {\tilde f}^m}\Biggr).\:
\label{vectfield}
\end{eqnarray}
Here $\check F^C_B\equiv F^C_B(F(Q^{\ast},a),a^{-1})$ is an inverse matrix to the matrix $F^A_B(Q^{\ast},a)$,
${\chi}^{\mu}_C\equiv \frac{\partial {\chi}^{\mu}(Q)}{\partial Q^C}|_{Q=Q^{\ast}}$, $({\Phi}^{-1})^{\beta}_{\mu}\equiv({\Phi}^{-1})^{\beta}_{\mu}(Q^{\ast})$ -- the matrix which is inverse to the Faddeev--Popov matrix:
\[
 ({\Phi})^{\beta}_{\mu}(Q)=K^A_{\mu}(Q)\frac{\partial {\chi}^{\beta}(Q)}{\partial Q^A},
\]
the matrix $\bar{v}^{\alpha}_{\beta}(a)$ is inverse of the matrix $\bar{u}^{\alpha}_{\beta}(a)$.\footnote{$\bar{u}^{\alpha}_{\beta}(a)$ (and ${u}^{\alpha}_{\beta}(a)$) are the coordinate representations of the auxiliary functions given on the group $\mathcal G$.}

The operator $N^A_C$, defined  as
\[
 N^A_C(Q)=\delta^A_C-K^A_{\alpha}(Q)({\Phi}^{-1})^{\alpha}_{\mu}(Q){\chi}^{\mu}_C(Q), 
\]
 is the projection operator ($N^A_BN^B_C=N^A_C$) onto the subspace which is orthogonal to the Killing vector field $ K^A_{\alpha}(Q)\frac{\partial}{\partial Q^A}$. $N^A_C(Q^{\ast})$ is the restriction of $N^A_C(Q)$ to the submanifold $ \Sigma $:
\[
 N^A_C(Q^{\ast})\equiv N^A_C(F(Q^{\ast},e))\;\;\;N^A_C(Q^{\ast})=F^B_C(Q^{\ast},a)N^M_B(F(Q^{\ast},a))\check F_M^A(Q^{\ast},a)
\]
$e$ is the unity element of the group.

Thus, the metric (\ref{metr_orig}) of the original manifold $\mathcal P \times V$ in a new coordinate basis is given by
\begin{equation}
\displaystyle
{\tilde G}_{\cal A\cal B}(Q{}^{\ast},\tilde f,a)=
\left(
\begin{array}{ccc}
 G_{CD}(P_{\bot})^C_A (P_{\bot})^D_B & 0 & G_{CD}(P_{\bot})^C_AK^D_{\nu}\bar u^{\nu}_{\alpha}\\
 0 & G_{mn} & G_{mp}K^p_{\nu}\bar u^{\nu}_{\alpha}\\
G_{CD}(P_{\bot})^C_AK^D_{\mu}\bar u^{\mu}_{\beta} & G_{np}K^p_{\nu}\bar u^{\nu}_{\beta} & d_{\mu\nu}\bar u^{\mu}_{\alpha}\bar u^{\nu}_{\beta}\\
\end{array}
\right)
\label{metric2c}
\end{equation}
where $G_{CD}(Q{}^{\ast})\equiv G_{CD}(F(Q{}^{\ast},e))$:
\[
 G_{CD}(Q{}^{\ast})=F^M_C(Q{}^{\ast},a)F^N_D(Q{}^{\ast},a)G_{MN}(F(Q{}^{\ast},a)), 
\]
  $(P_\bot)^A_B$ is  the projection operator on the tangent plane to the submanifold $\Sigma$. It is given by
\[
 (P_\bot)^A_B=\delta^A_B-\chi ^{\alpha}_{B}\,(\chi \chi ^{\top})^{-1}{}^{\beta}_{\alpha}\,(\chi ^{\top})^A_{\beta}, 
\]
$(\chi ^{\top})^A_{\beta}$ is a transposed matrix to the matrix $\chi ^{\nu}_B$:
\[
 (\chi ^{\top})^A_{\mu}=G^{AB}{\gamma}_{\mu\nu}\chi ^{\nu}_B\;\;\;\;{\gamma}_{\mu\nu}=K^A_{\mu}G_{AB}K^B_{\nu}.
\]
($(P_\bot)^A_B$ has the following properties:
 $(P_\bot)^A_BN^C_A=(P_\bot)^C_B,N^A_B(P_\bot)^C_A=N^C_B.$)

 $d_{\mu\nu}(Q{}^{\ast},\tilde f)\bar u^{\mu}_{\alpha}(a)\bar u^{\nu}_{\beta}(a)$ in (\ref{metric2c}) is 
the metric on  $\mathcal G$--orbit through the point $(p,v)$: 
\begin{eqnarray*}
 d_{\mu\nu}(Q{}^{\ast},\tilde f)&=&K^A_{\mu}(Q{}^{\ast})G_{AB}(Q{}^{\ast})K^B_{\nu}(Q{}^{\ast})+K^m_{\mu}(\tilde f)G_{mn}K^n_{\nu}(\tilde f)
\nonumber\\
&\equiv&\gamma_{\mu \nu}(Q^{\ast})+\gamma'_{\mu \nu}(\tilde f).
\end{eqnarray*}

In our works \cite {Stor_lagr_poinc_1,Stor_lagr_poinc_2}, the Lagrange-Poincar\'e  equations equations were derived using the so-called the horizontal lift basis  on the total space of the principal fiber bundle. 
The new basis consists of the horizonal and vertical vector fields and can be determined by using the ``mechanical connection'' which 
exists \cite{AbrMarsd} in  case of the reduction of  mechanical systems with a symmetry.

In the principal fiber bundle ${\rm P}(\mathcal P\times_{\mathcal G}V,\mathcal G)$,  in coordinates $(Q^{\ast}{}^A,\tilde f^n,a^{\alpha})$, the connection  $\hat\omega=\hat\omega ^{\alpha}\otimes\lambda_{\alpha}$ ($\{\lambda_{\alpha}\}$ is the basis in the Lie algebra of the group Lie $\mathcal G$) is given by the following expression:
\[
 \hat\omega ^{\alpha}=\bar {\rho}^{\alpha}_{\alpha '}(a)\biggl(d^{\alpha '\mu}K^D_{\mu}(Q{}^{\ast})G_{DA}(Q{}^{\ast})dQ{}^{\ast  A}+d^{\alpha '\mu}K^q_{\mu}(\tilde f)G_{qn}d{\tilde f}^n\biggr)+u^{\alpha}_{\beta}(a)da^{\alpha},
\]
where   $d^{\alpha '\mu}= d^{\alpha '\mu}(Q{}^{\ast},\tilde f)$. And the  matrix  $\bar {\rho}^{\alpha}_{\alpha '}(a)$ 
 is inverse to the matrix ${\rho}_{\alpha}^{\beta}={\bar u}^{\alpha}_{\nu}v^{\nu}_{\beta}$  of the adjoint representation of the group $\cal G$,

In terms of the (``gauge'') potentials $\mathscr A^{\alpha }_B$ and $\mathscr A^{\alpha '}_m$, together with  a new notation: $\tilde\mathscr A^{\alpha }_B=\bar {\rho}^{\alpha}_{\alpha '}(a)\mathscr A^{\alpha '}_B(Q{}^{\ast},\tilde f)$, the connection can be rewritten as 
\begin{equation}
\hat\omega ^{\alpha}= 
\tilde\mathscr A^{\alpha }_B(Q{}^{\ast},\tilde f,a)dQ{}^{\ast  B}+\tilde\mathscr A^{\alpha }_m(Q{}^{\ast},\tilde f,a)d{\tilde f}^m+u^{\alpha}_{\beta}(a)da^{\alpha},
\label{connect_Q_star}
\end{equation}
or  using the  condensed notations of  indices like
\[
 \hat\omega ^{\alpha}= 
\tilde\mathscr A^{\alpha '}_{\tilde B}(Q{}^{\ast},\tilde f,a)dQ{}^{\ast  \tilde B}+u^{\alpha}_{\beta}(a)da^{\alpha}.
\]

In coordinates $(Q^{\ast}{}^A,\tilde f^n,a^{\alpha})$, the horizontal lift basis $(H_A,H_p,L_{\alpha})$ is given by  the vector fields 
\[
H_M(Q^{\ast},\tilde f,a)= \Bigl[N^T_M\Bigl(\frac{\partial}{\partial Q^{\ast T}}-\tilde \mathscr A^{\alpha }_T L_{\alpha}\Bigl)+N^m_M\Bigl(\frac{\partial}{\partial {\tilde f}^m}-\tilde \mathscr A^{\alpha }_mL_{\alpha}\Bigl)\Bigr],
\]
\[
 H_m(Q^{\ast},\tilde f,a)=\Bigl( \frac{\partial}{\partial {\tilde f}^m}-\tilde \mathscr A^{\alpha }_mL_{\alpha}\Bigl),
\]
and also by the left-invariant vector field    $L_{\alpha}=v^{\nu}_{\alpha}(a)\frac{\partial}{\partial a^{\nu}}$  which  is obtained from the Killing vector field $K_{\alpha}(Q)$. 
Note that $L_{\alpha}$ commutes with the horizontal vector fields $H_A$ and $H_p$.

In the definition of $H_M$, new components of the projection operator 
$$N^{\tilde A}_{\tilde C}=(N^A_C,N^A_m,N^m_A,N^m_p),$$ 
 were used. They are
\[
N^A_m=0,\;\;\;  N^m_A=-K^m_{\alpha}({\Phi}^{-1})^{\alpha}_{\mu}\,{\chi}^{\mu}_A=-K^m_{\alpha}{\Lambda}^{\alpha}_A,\;\;\;  N^m_p={\delta}^m_p.
\]
The operator $N^{\tilde A}_{\tilde B}$ satisfy the following property: $N^{\tilde A}_{\tilde B} N^{\tilde B}_{\tilde C}=N^{\tilde A}_{\tilde C}$.

In a new coordinate basis $(H_A,H_p,L_{\alpha})$,  the  metric tensor (\ref{metric2c})  is represented as
\begin{equation}
\displaystyle
{\check G}_{\cal A\cal B}(Q^{\ast},\tilde f,a)=
\left(
\begin{array}{ccc}
{\tilde G}^{\rm H}_{AB} & {\tilde G}^{\rm H}_{Am} & 0\\
{\tilde G}^{\rm H}_{nB} & {\tilde G}^{\rm H}_{nm} & 0\\
0 & 0 &\tilde{d}_{\alpha \beta }  \\
\end{array}
\right)\equiv\left( \begin{array}{cc}
{\tilde G}^{\rm H}_{\tilde A \tilde B}  & 0 \\
0 & \tilde{d}_{\alpha \beta }  \\
\end{array}
\right),
\label{metric2cc}
\end{equation}
where $\tilde{d}_{\alpha \beta }=\rho^{\alpha'}_{\alpha}\rho^{\beta'}_{\beta}d_{\alpha' \beta' }$. 
The  components of the ``horizontal metric'' ${\tilde G}^{\rm H}_{\tilde A \tilde B}$ depending on $(Q^{\ast}{}^A,\tilde f ^m)$ are defined as follows:
\begin{eqnarray*}
&&{\tilde G}^{\rm H}_{AB}={\Pi}^{\tilde A}_{A}\,{\Pi}^{\tilde B}_B \,G_{\tilde A\tilde B}=G_{AB}-G_{AD}K^{D}_{\alpha}d^{\alpha \beta}K^R_{\beta}\,G_{RB},
\nonumber
{}\\
&&{\tilde G}^{\rm H}_{Am}=-G_{AB}K^{B}_{\alpha}\,d^{\alpha \beta}K^p_{\beta}G_{pm},
\nonumber
{}\\
&&{\tilde G}^{\rm H}_{mA}=-G_{mq}K^{q}_{\mu}\,d^{\mu \nu}K^D_{\nu}G_{DA},
\nonumber
{}\\
&&{\tilde G}^{\rm H}_{mn}= G_{mn}-G_{mr}K^r_{\alpha}d^{\alpha \beta}K_{\beta}^pG_{pn}.
\nonumber\\
\end{eqnarray*}

In the coordinate basis $(H_A,H_p,L_{\alpha})$, the original Lagrangian $\mathcal L$
has the following representation:
\begin{equation}
 {\hat{\mathcal L}}=\frac12\,({\tilde G}^{\rm H}_{AB}\, {\omega}^A {\omega}^B +{\tilde G}^{\rm H}_{Ap}\, {\omega}^A {\omega}^p+{\tilde G}^{\rm H}_{pA}\, {\omega}^p {\omega}^A+{\tilde G}^{\rm H}_{pq}\, {\omega}^p {\omega}^q+ {\tilde{d}}_{\mu \nu} {\omega}^{\mu} {\omega}^{\nu})-V,
\label{lagrang_3}
\end{equation}
where  the new time-dependent variables ${\omega}^A,{\omega}^p$ and ${\omega}^{\alpha}$, which are associated with  velocities, are given by
\begin{eqnarray} 
&&\omega ^A=(P_{\bot})^A_B\, \frac{dQ^{\ast}{}^B}{dt}=\frac{dQ^{\ast}{}^A}{dt},\;\;\;\;
\omega ^p=\frac{d\tilde f^p}{dt}\nonumber\\
&&{\omega}^{\alpha}=u^{\alpha}_{\mu}\frac{da^{\mu}}{dt}+{\tilde {\mathscr A}}^{\alpha}_E\,\frac{dQ^{\ast}{}^E}{dt}+{\tilde {\mathscr A}}^{\alpha}_m\, \frac{d\tilde f^m}{dt}.
\label{veloc_omega}
\end{eqnarray}

The Lagrangian (\ref{lagrang_3}) was used in \cite{Stor_lagr_poinc_1, Stor_lagr_poinc_2} for derivation of the Lagrange-Poincar\'{e} equations. This was done  using the Poincar\'{e} variational principle. The following equations were obtained:

\begin{eqnarray*}
&&\!\!\!\!\!\!\!\!\!\!N^A_B\frac{d{\omega}^B}{dt}+N^A_R\,{}^{\rm  H}\tilde {\Gamma}^R_{\tilde B\tilde M}{\omega}^{\tilde B} {\omega}^{\tilde M} 
\nonumber\\
&&\qquad {} +G^{EF}N^A_EN^{\tilde R}_F\Bigl[{ \mathscr F}^{\alpha}_{\tilde Q\tilde R}  {\omega}^{\tilde Q} p_{\alpha} +\frac12({\mathscr D}_{\tilde R} {d}^{\kappa\sigma}) p_{\kappa}p_{\sigma}+V_{,\tilde R}\Bigr]=0,                                 
{}\\
\abovedisplayskip=0pt
&&\!\!\!\!\!\!\!\!\!\!N^r_B\frac{d{\omega}^B}{dt}+\frac{d{\omega}^r}{dt}+N^r_{\tilde R}\,{}^{\rm  H}\tilde {\Gamma}^{\tilde R}_{\tilde A\tilde B}{\omega}^{\tilde A} {\omega}^{\tilde B}+
G^{EF}N^r_FN^{\tilde R}_E\Bigl[{ \mathscr F}^{\alpha}_{\tilde Q\tilde R}  {\omega}^{\tilde Q} p_{\alpha}+
\nonumber\\
&&\!\!\!\!\!\!\!\!\!\!\frac12({\mathscr D}_{\tilde R} {d}^{\kappa\sigma}) p_{\kappa}p_{\sigma}+V_{,\tilde R}\Bigr]+
G^{rm}\Bigl[{ \mathscr F}^{\alpha}_{\tilde Q m}  {\omega}^{\tilde Q} p_{\alpha}+\frac12({\mathscr D}_{m} {d}^{\kappa\sigma}) p_{\kappa}p_{\sigma}+V_{,m}\Bigr]=0.
{}\\
&&\!\!\!\!\!\!\!\!\!\!\frac{d p_{\beta}}{dt}+ c^{\nu}_{\mu \beta}d^{\mu \sigma}p_{\sigma}p_{\nu}-c^{\nu}_{\sigma \beta}\mathscr A^{\sigma}_{\tilde E}\omega ^{\tilde E}p_{\nu}=0.
\end{eqnarray*}
(Here  the condensed notation for indices is used: according to which the sum over the repeated index $\tilde R$ means the summation over $R$ and $r$.)

In these equations $p_\sigma={\gamma}_{\alpha\sigma}{\rho}^{\alpha}_{\epsilon}{\omega}^{\epsilon}$,
the curvature tensor ${\mathscr F}^{\alpha}_{SP}$ of the connection ${{\mathscr A}^{\alpha}_P}$ is given by
${\mathscr F}^{\alpha}_{SP}= {\mathscr A}^{\alpha}_{P,S}- 
{\mathscr A}^{\alpha}_{S,P}
+c^{\alpha}_{\nu\sigma}\, {\mathscr A}^{\nu}_S\,
{\mathscr A}^{\sigma}_P$.
The tensors ${\mathscr F}^{\alpha}_{Ep}$ and ${\mathscr F}^{\alpha}_{pm}$ are defined in a similar way.

The covariant derivative ${\mathscr D}_{R}( {d}^{\kappa\sigma}(Q^{\ast},\tilde f))$ are given by 
$$\displaystyle{\mathscr D}_{R} {d}^{\kappa\sigma}=
\partial_{Q^{\ast  R}}
{d}^{\kappa\sigma} +\mathscr A^{\alpha}_R c^{\kappa}_{\alpha \nu}d^{\nu \sigma}+\mathscr A^{\alpha}_R c^{\sigma}_{\alpha \nu}d^{\nu \kappa}. $$

The Christoffel symbols ${}^{ \mathrm  H}{\tilde \Gamma}_{BM}^{\tilde R}$, ${}^{ \mathrm  H}{\tilde \Gamma}_{qB}^{\tilde R}$ and ${}^{ \mathrm  H}{\tilde \Gamma}_{pq}^{\tilde R}$ for the horizontal (degenerate) metric ${\tilde G}^H_{\tilde R \tilde T}$ are  defined 
by means   of the  equalities:
\[
 {}^{ \mathrm  H}{\tilde \Gamma}_{BM\tilde T}={\tilde G}^H_{\tilde R \tilde T}{}^{ \mathrm  H}{\tilde \Gamma}_{BM}^{\tilde R},\;\;\;{}^{ \mathrm  H}{\tilde \Gamma}_{qB\tilde T}={\tilde G}^H_{\tilde R \tilde T}{}^{ \mathrm  H}{\tilde \Gamma}_{qB}^{\tilde R}\;\;\;{\rm and}\;\;{}^{ \mathrm  H}{\tilde \Gamma}_{pq\tilde T}={\tilde G}^H_{\tilde R \tilde T}{}^{ \mathrm  H}{\tilde \Gamma}_{pq}^{\tilde R},
\]
where 
\[
{}^{ \mathrm  H}{\tilde \Gamma}_{BMD}\equiv\frac12({\tilde G}^{\rm H}_{BD,M}+{\tilde G}^{\rm H}_{MD,B}-{\tilde G}^{\rm H}_{BM,D}).
\]
And ${}^{ \mathrm  H}{\tilde \Gamma}_{qBT}$ and ${}^{ \mathrm  H}{\tilde \Gamma}_{pqT}$ have an analogous  definitions.

Taking into account  the following  properties: 
\[
K^{\tilde E}_{\alpha}{ }^{\;\rm H}{\Gamma}_{A\,C\tilde E}=0,\quad K^{\tilde R}_{\beta}\,{ \mathscr F}^{\alpha}_{\tilde Q\tilde R}=0,\quad N^{\tilde T}_F\,{ }^{\;\rm H}{\Gamma}_{B\,M\tilde T}={ }^{\;\rm H}{\Gamma}_{BMF}, 
\]
\[
 N^{\tilde T}_F\,{ }^{\;\rm H}{\Gamma}_{B\,M\tilde T}={ }^{\;\rm H}{\Gamma}_{BMF},\quad
 N^{\tilde T}_F\,{ }^{\;\rm H}{\Gamma}_{q\,B\tilde T}={ }^{\;\rm H}{\Gamma}_{qBF},\quad
 N^{\tilde T}_F\,{ }^{\;\rm H}{\Gamma}_{q\,p\tilde T}={ }^{\;\rm H}{\Gamma}_{qpF},
\]
\[
  K^{\tilde R}_{\varepsilon}{\mathscr D}_{\tilde R} ({d}_{\alpha\beta})=0,\quad
 N^{\tilde R}_{F}{\mathscr D}_{\tilde R} ({d}_{\mu\nu})={\mathscr D}_{F} ({d}_{\mu\nu}),
\]
and  the  invariance of the potential $V(Q^{\ast},\tilde f)$ under the action of the group $\mathcal G$,  this means that $N^{\tilde R}_FV_{\!,\tilde R}=V_{\!, F}$,
we can rewrite the Lagrange-Poincar\'{e} equations in the following way:
\begin{eqnarray}
&&\!\!\!\!\!\!\!\!\!\!N^B_A\Bigl(\frac{d{\omega}^A}{dt}+G^{AR}\,{}^{\rm  H}\tilde {\Gamma}_{\tilde B\tilde MR}\,{\omega}^{\tilde B} {\omega}^{\tilde M} 
\nonumber\\
&&\qquad {} +G^{AR}\Bigl[{ \mathscr F}^{\alpha}_{\tilde Q\, R}  {\omega}^{\tilde Q} p_{\alpha} +\frac12({\mathscr D}_{ R} {d}^{\kappa\sigma}) p_{\kappa}p_{\sigma}+V_{,R}\Bigr]\Bigl)=0.                                 
\label{itog_hor_1}\\
&&{}\nonumber\\
&&\!\!\!\!\!\!\!\!\!\!\!\!\!N^r_A\Bigl[\frac{d{\omega}^A}{dt}+G^{AR}\Bigl({}^{\rm  H}\tilde {\Gamma}_{\tilde B\tilde M R}{\omega}^{\tilde B} {\omega}^{\tilde M}
   + { \mathscr F}^{\alpha}_{\tilde Q R}  {\omega}^{\tilde Q} p_{\alpha}+\frac12({\mathscr D}_{} {d}^{\kappa\sigma}) p_{\kappa}p_{\sigma}+V_{, R}\Bigr)\Bigr]\!+
\nonumber\\
&&\!\!\!\!\!\!\!\!\!\!\!\!\!\frac{d{\omega}^r}{dt}+G^{rm}\Bigl({}^{\rm  H}\tilde {\Gamma}_{\tilde B\tilde M m}{\omega}^{\tilde A} {\omega}^{\tilde B}
 +
{ \mathscr F}^{\alpha}_{\tilde Q m}  {\omega}^{\tilde Q} p_{\alpha}+\frac12({\mathscr D}_{m} {d}^{\kappa\sigma}) p_{\kappa}p_{\sigma}+V_{,m}\Bigr)=0.
\label{itog_hor_2}\\
&&{}\nonumber\\
&&\!\!\!\!\!\!\!\!\!\!\!\!\!\frac{d p_{\beta}}{dt}+ c^{\nu}_{\mu \beta}d^{\mu \sigma}p_{\sigma}p_{\nu}-c^{\nu}_{\sigma \beta}\mathscr A^{\sigma}_{\tilde E}\omega ^{\tilde E}p_{\nu}=0.
\label{itog_vert}
\end{eqnarray}

These equations will be used for derivation of  the Lagrange-Poincar\'{e} equations
in gauge theories.

\section{Adapted coordinates in configuration space of the  gauge system with interaction}
Our aim is to extend the methods we have used for the finite-dimensional mechanical system with symmetry to the gauge system which describe the dynamics of Yang-Mills field interacting with the scalar field. The standard relativistically invariant Lagrangian for this system is singular
(we can not determine the Hamiltonian using the Legendre transformation) in contrast to what we had for the model mechanical system. The problem is related to presence of the redundent variable $A^a_0$ in the Lagrangian.  
 Therefore, by setting $A^a_0=0$ in the  Lagrangian we obtain the Lagrangian which is free of this problem. Note, that the same can be performed  by suitable gauge transformation

So, the Lagrangian (Lagrange density) we will consider is the following
\begin{eqnarray}
\mathcal L&=& -\frac{1}{2g_0^2}k_{\alpha\beta}(\partial_0A_i^{\alpha})(\partial_0A^{i\beta})+\frac12G_{ab}(\partial_0f^a)(\partial_0f^{b})\nonumber\\
&&-\frac{1}{4g_0^2}k_{\alpha\beta}F^{\alpha}_{ij}F^{\beta ij}+\frac12G_{ab}(\nabla_if^a)(\nabla^if^b)-V_0(A,f) .
\label{lagr_0}
\end{eqnarray}
Here $k_{\alpha \beta}=c^{\tau}_{\mu \alpha}c^{\mu}_{\tau \beta}$ is the Cartan--Killing metric on the group $\mathcal G$, $V_0$ is some gauge-invarint potential. 
$g_0$ is a coupling constant.\footnote{Further, in the formulas, we omit the coupling constant  $g_0$, absorbing it in $k_{\alpha \beta}$ since
in the final expressions, the coupling constant can be easily restored.}

The covariant derivative $\nabla _i$ is defined as follows:
$$(\nabla f)^a_i(\bar x,t)=(\delta^a_b\partial_i(\bar x)-(\bar J_{\alpha})^a_bA^{\alpha}_i(\bar x,t)\,)f^b(\bar x,t),$$
where ${\bar J}_{\alpha}$ are the generators of the representation ${\bar D}^n_m(a)$ which acts (on the right) in the vector space $V$: 
$\hat f^n= {\bar D}^n_m(a)f^m$, ${\bar D}^n_m(\Phi(g,h))={\bar D}^m_p(h){\bar D}^p_n(g)$.
The generators satisfy the  following commutation relation 
$[{\bar J}_{\alpha},{\bar J}_{\beta}]={\bar c}^{\gamma}_{\alpha \beta}{\bar J}_{\gamma}$, where the structure constants
${\bar c}^{\gamma}_{\alpha \beta}=-{c}^{\gamma}_{\alpha \beta}$.

The  Lagrangian (\ref{lagr_0}) is  invariant under time-independent  gauge transformations  of the gauge potentials and scalar fields:
:
\begin{eqnarray*}
{\tilde A}^{\alpha}_i({\mathbf x})&=&{\rho}^{\alpha}_{\beta}(g^{-1}({\mathbf x}))
{ A}^{\beta}_i({\mathbf x})+u^{\alpha}_{\mu}(g({\mathbf x}))
\frac{\partial g^{\mu}({\mathbf x})}
{\partial {\mathbf x}^i}\,,
\nonumber\\
 \tilde f^a(\mathbf x,t)&=& {\bar D}^a_b(g(\mathbf x)) f ^b(\mathbf x,t).
\end{eqnarray*}


The obtained Lagrangian looks as if it represents the motion of two ``particle'' in the product space $\mathscr P\times \mathscr V$ in the potential
\[
V[A,f]=\int d^3x\Bigl[\frac{1}{2}\,
k_{\alpha \beta}\,
F^{\alpha}_{ij}({\mathbf x})\,F^{\beta \;ij}({\mathbf x})-\frac12G_{ab}(\nabla f)^a_i(\mathbf x)(\nabla f)^{b\,i}(\mathbf x)+V_0\Bigr ].
\]

 One of the space, $\mathscr P$, is  an infinite-dimensional Riemannian manifold. The gauge fields $A_i^a$ can be regarded as points of this manifold. And the other space, $\mathscr V$, is the space of functions with the values in the vector space $\mathcal V$. Also, we are given  an action of the group, the group of the gauge transformations, on the product space. 
This is analogous to what we have in reduction problem for dynamical system with symmetry in mechanics, which was  considered in the previous section. Here we are interested in description of internal dynamics given on the gauge orbit space.

The reduction theory for the gauge-invariant dynamical systems follows from the result obtained  in \cite{Narasimhan,                                                                                                                                                                                                                                                                                                                                                                                                                                                                                                                                                                                                                                                                                                                           Mitter, Daniel-Viallet, Mitter-Viallet, Singer_2, Babelon-Viallet}, and in other works, where 
 the geometric approach to the gauge fields was developed.

First of all, the gauge fields $ A^a_{\mu} (x)$  are regarded as 
coordinate representations of connections
defined on the principal fiber bundle $P(M,\mathcal G)$ over the compact manifold $M$.\footnote{Using compact manifolds 
 needed
to ensure the boundedness of the action functional \cite{Daniel-Viallet, Postnikov}.}
Then, in order to have  a smooth free and proper action of the gauge group  on the space of connections $\mathscr P$, one must consider the irreducible connections. (The isotropy subgroup of these connections coincides with  ${ {\mathscr Z}(G)}$, the center of gauge group ${\mathscr G}$.)
The gauge transformation group must be group $\tilde{\mathscr G}={\mathscr G}/{\mathscr Z}(G)$. Moreover, 
the connections and the gauge transformation functions must belong to  classes of Sobolev functions $H_k$ and $H_{k+1}$,  respectively, with $k\geq3$ \cite{Narasimhan, Mitter-Viallet}.
Only in this case one  leads to the principal fiber bundle defined by $\pi :\mathscr P\to  \mathscr P/ \tilde{\mathscr G}=\mathscr M$.

The function space $\mathscr V$ of the matter fields $f^b(\mathbf x,t)$ consists of the sections of the associated bundle $\Gamma (\mathcal P\times_{\mathcal G}\mathcal V)$. (These sections also must be from $H_k$.)

In our case,  $ \mathscr P\times \mathscr V$ is  the  original configuration space of the  gauge  system with the Lgrangian (\ref{lagr_0}),
and the gauge orbit space $\mathscr P\times_{\mathscr G} \mathscr V$, the base of the principal fiber bundle $\pi': \mathscr P\times \mathscr V\to (\mathscr P\times \mathscr V)/\tilde{\mathscr G}=\mathscr P\times_{\tilde\mathscr G} \mathscr V$, is the configuration space of the physically observable quantities.

From the quadratic part of the Lagrangian (\ref{lagr_0}) it follows that the Riemannian metric of the original configuration space is flat. It can be presented as follows:
\[
ds^2=G_{(\alpha , i, x)(\beta,j,y)}\delta A^{(\alpha ,i,x)}\,
\delta A^{(\beta ,j,y)}+G_{(a, x)(b,y)}\,\delta f^{(a,x)}
\delta f^{(b,y)},
\]
where 
\[
 G\biggl(\frac{\delta}{\delta A^{(\alpha,i,x)}},\frac{\delta}{\delta A^{(\beta,j,y)}}\biggr)=G_{(\alpha,i,x)(\beta,j,y)}={k}_{\alpha\,\beta}\,
{\delta}_{i\,j}\,{\delta}^3({\mathbf x}-{\mathbf y})
\]
is the metric on $\mathscr P$ and the metric on $\mathscr V$ is
\[
 G\biggl(\frac{\delta}{\delta f^{(m,x)}},\frac{\delta}{\delta f^{(n,y)}}\biggr)=G_{(m,x)(n,y)}=G_{mn}\delta^3(\mathbf x-\mathbf y).
\]
In these formulae we have used the extended notation for indices by which  $A^{(\alpha,i,x)}\equiv A^{\alpha i}(\mathbf x)$ and  $f^{(m,x)}\equiv f^m(\mathbf x)$. Note that the use of such notations helps in the generalization of formulas obtained in the finite-dimensional case to the corresponding formulas in field theories.

From the gauge invariance of the Lagrangian (and the metric) it follows that the Killing vectors of the original metric are
\[
K_{(\alpha,y)}=K^{(\mu, i, x)}_{\;\;\;\;\;\;(\alpha,y)}\frac{\delta}
{\delta A^{(\mu ,i,x)}}+K^{(b, x)}_{\;\;\;\;(\alpha,y)}\frac{\delta}
{\delta f^{(b,x)}},
\]
where components of this vector field are given by
\[
K^{(\mu, i, x)}_{\;\;\;\;\;\;(\alpha,y)}(A)=
\left[\left({\delta}^{\;\mu}_{\alpha}{\partial}^i({\mathbf x})
+c^{\mu}_{\tilde \nu \alpha}A^{\tilde \nu i}({\mathbf x})
\right){\delta}^3 ({\mathbf x}-{\mathbf y})\right]
\equiv \left[{\mathcal D}^{\mu i}_{\;\;\alpha}(A({\mathbf x}))
\,{\delta}^3({\mathbf x}-{\mathbf y})\right]
\]
(here ${\partial}_i({\mathbf x})$ is a partial derivative 
with respect to $x^i$), and
\[
 K^{(b,x)}_{\;\;\;\;(\alpha, y)}(f)=(\bar J_{\alpha})^b_cf^c(\mathbf x)\delta ^3(\mathbf x-\mathbf y).
\]

We can determine the coordinates in the principal bundle for the gauge system under study just  the same way as was done for a mechanical system with symmetry in a finite-dimensional space.  ?The local sections $\Sigma$ of the principal fiber bundle $\rm P(\mathscr M, \tilde\mathscr G)$, which are necessary for determination of the bundle coordinates in the total space $\mathscr P\times\mathscr V$ of the bundle $\pi'$,  will be defined by means of the Coulomb gauge condition (or the Coulomb gauge): $\partial _i A^{\alpha i}=0$. The  gauge potentials that will satisfy this equation (dependent variables) will be denoted by $A^{\ast}_i{}^{\alpha}$. 
 Note that dependent variables are typically used when quantizing gauge fields \cite{Babelon-Viallet, Creutz, Gawedzki, Huffel-Kelnhofer, Kunstatter, Falck}.

As was shown in previous section, for  transition from the original coordinate $(A^{\alpha}_i,f^a)$ given on $\mathscr P\times\mathscr V$ to the adapted coordinates $(A^{\ast}_i{}^{\alpha},\tilde f^b, a^{\mu})$ of the principal fiber bundle it is requiered that the group coordinates $a^{\alpha}(A)$ of the ``point'' $A$ should be known. In mechanics, they are obtained as a solution of the equation (\ref{a_chi}):      $\chi^{\beta}(F^A(Q, a^{-1}(Q)))=0.$ For the Coulomb gauge, this equation is as follows:
\[
\partial^i({\mathbf x})\left[\, {\rho}^{\alpha}_
{\;\beta}(a({\mathbf x}))\,
A^{\beta}_{\,i}({\mathbf x})-{\rho}^{\alpha}_{\;\nu}(a({\mathbf x}))\,
u^{\nu}_{\;\sigma}(a({\mathbf x}))\,\frac{\partial 
a^{\sigma}({\mathbf x})}
{\partial {\mathbf x}^i}\,\,\right]=0\,.
\]

Then, the  coordinates $Q^{\ast}$ of the corresponding point 
on a submanifold $\Sigma $ are determined by the corresponding group transformation:
$$Q^{\ast}{}^A=F^A(Q,a^{-1}(Q)).$$ In gauge theories, we have  the following gauge transformation:
\[
A^{\alpha}_i({\mathbf x})={\rho}^{\alpha}_{\beta}(a^{-1}({\mathbf x}))
{ A^{\ast}}^{\beta}_i({\mathbf x})+u^{\alpha}_{\mu}(a({\mathbf
x}))\frac{\partial
a^{\mu}({\mathbf x})}
{\partial {\mathbf x}^i}\,.
\]
With the obtained $a^{\alpha}({\mathbf x})$, $f^a$ is expressed in terms of $\tilde f^a$ as follows:  $f^a(\mathbf x)={\bar D}^a_b(a({\mathbf x}))\tilde f^b(\mathbf x)$. Thus, the initial coordinates $(A^{\alpha}_i(\mathbf x), f^a(\mathbf x))$ on  $\mathscr P\times \mathscr V$ are transformed into adapted bundle coordinates $(A^{\ast}{}^{\alpha}_i({\mathbf x}), \tilde f^b(\mathbf x), a^{\alpha}(\mathbf x))$. 

To obtain a new coordinate representation of the original Riemannian metric, we must transform the coordinate vector fields. 
The ``vector fields'' transformation formula  is  a strightforward
generalization of the corresponding formula from the finite-dimensional case:
\begin{eqnarray}
&&\frac{\delta}{\delta A^{(\alpha,i,x)}}=
\check F^{(\mu, k,u)}_{\;\;\;\;\;\;(\alpha,i,x)}\,\Bigl(
N^{(\nu,p,v)}_{\;\;\;\;\;\;(\mu, k,u)}(A^{\ast})\,
\frac{\delta}{\delta A^{\ast}{}^{(\nu,p,v)}}+N^{(m,y)}_{(\mu,k,u)}\frac{\delta}{\delta f^{(m,y)}}\Bigr.
\nonumber\\
&&\;\;\;\;
\Bigl.+{\chi}^{(\mu',v)}_{\;\;\;(\mu, k,u)}(A^{\ast})\,
\bigl(\Phi ^{-1}\bigr)^{(\beta,u')}_{\;\;\;(\mu',v)}(A^{\ast})\,
{\bar v}^{(\sigma,p)}_{\;\;(\beta,u')}(a)\,
\frac{\delta}{\delta a^{(\sigma,p)}}\Bigr)\,,
\label{5}
\end{eqnarray}
where we have denoted by $\check F$ the matrix 
which is inverse to the matrix  
$F^{(\mu, k,u)}_{\;\;\;\;\;\;(\alpha,i,x)}$  defined as follows
\[
F^{(\alpha,i,x)}_{\;\;\;\;\;\;(\beta,j,y)}[A,a]=
\frac{\delta {\tilde A}^{(\alpha,i,x)}}
{\delta A^{(\beta,j,y)}}=
{\rho}^{\alpha}_{\beta}(g^{-1}({\mathbf x}))\, \delta^i_{\,j}\,
{\delta}^3  ({\mathbf x}-{\mathbf y})\,.
\]
$\check F$ satisfies the relation:
\[
F^{(\alpha,i,x)}_{\;\;\;\;\;\;(\beta, j,y)}\,
\check F^{(\beta, j,y)}_{\;\;\;\;\;\;(\epsilon, k,z)}=
{\delta}^{\alpha}_{\,\epsilon}\,{\delta}^i_{\,k}\,
{\delta}^3({\mathbf x}-{\mathbf z})\,.
\]
Also, we have 
\[
 \frac{\delta}{\delta f^{(n,x)}}=D^m_n(a(\mathbf x))\frac{\delta}{\delta \tilde f^m(\mathbf x)}.
\]

In formula (\ref{5}), by $N^{(\nu,p,v)}_{\;\;\;\;\;\;(\mu, k,u)}$,
which is equal to
\[
N^{(\alpha,i,x)}_{\;\;\;\;\;\;(\beta,j,y)}=
{\delta}^{(\alpha,i,x)}_{\;\;\;\;\;\;(\beta,j,y)}
-K^{(\alpha,i,x)}_{\;\;\;\;\;\;(\mu,z)}
\bigl({\Phi}^{-1}\bigr)^{(\mu,z)}_{\,\,\;\;(\nu,u)}
{\chi}^{(\nu,u)}_{\;\;\;\;(\beta,j,y)}\,,
\]
we have denoted the projection operator 
onto the subspace which is orthogonal to the component of the Killing 
vector field $K_{(\alpha,y)}$ which is related to $\mathscr P$.

The projection operator $N^{(m,y)}_{(\mu,k,u)}$ is equal to  
\[
 N^{(m,y)}_{(\mu,k,u)}=-K^{(m,y)}_{\,\,\,(\alpha,z)}\bigl({\Phi}^{-1}\bigr)^{(\alpha,z)}_{\,\,\;\;(\beta,v)}
{\chi}^{(\beta,v)}_{\;\;\;\;(\mu,k,u)}\,. 
\]

The Faddeev--Popov matrix $\Phi $ is defined as follows
\[
{\Phi}^{(\nu,y)}_{\;\;(\mu,z)}\bigl[A\bigr]=
K^{(\alpha,i,x)}_{\;\;\;\;(\mu,z)}\;{\chi}^{(\nu,y)}_
{\;\;(\alpha,i,x)}\,.
\]
For the Coulomb gauge, we have
\[
{\chi}^{(\nu, y)}_{\;\;\;\;(\alpha,i,x)}=
{\delta}^{\nu}_{\;\alpha}
\left[\,{\partial }_i({\mathbf y})\;{\delta}^3({\mathbf y}-
{\mathbf x})\,\right]\,.
\]
Therefore, the matrix $\Phi $ (restricted to the gauge surface) is equal to
\[
{\Phi}^{(\nu,y)}_{\;\;\;\;(\mu,z)}[A^{\ast}]=
\left[\bigl({\delta}^{\nu}_{\mu}\;{\partial}^2({\mathbf y})
+c^{\nu}_{\sigma
\mu}{A^{\ast}}{}^{\sigma}_i({\mathbf y})\;
{\partial}^i({\mathbf y})\,\bigr)\;
{\delta}^3({\mathbf y}-{\mathbf z})\right]
\]
or
\[
{\Phi}^{(\nu,y)}_{\;\;\;\;(\mu,z)}[A^{\ast}]=
\left[\bigl(\,{\cal D}\,[A^{\ast}]\cdot\partial \,
\bigr)^{\nu}_{\mu}({\mathbf y})\;{\delta}^3({\mathbf y}-{\mathbf
z})\,\right]\,.
\]
An inverse matrix ${\Phi}^{-1}$ can be determined  by the equation
\[
{\Phi}^{(\nu,y)}_{\;\;\;\;(\mu,z)}\;
({\Phi}^{-1}){}^{(\mu,z)}_{\;\;\;\;(\sigma,u)}({\mathbf y},
{\mathbf u})=
{\delta}^{\nu}_{\sigma}\,{\delta}^3({\mathbf y}-{\mathbf u})\,.
\]
That is, it is the Green function for the Faddeev--Popov operator:
\[
\left[\partial_i({\mathbf y}){\cal
D}^{\nu\,i}_{\mu}[A({\mathbf y})]\,\right]\,
({\Phi}^{-1}){}^{(\mu,y)}_{\;\;\;\;(\sigma,u)}
({\mathbf y},{\mathbf u})=
{\delta}^{\nu}_{\sigma}\,{\delta}^3({\mathbf y}-{\mathbf u})\,.
\]
(The boundary conditions of this operator depend on a concrete choice
of a base manifold $M$.)
By a second group of variables, the Green function ${\Phi}^{-1}$  satisfies  the following equation:
\[
\left[-\tilde{\cal D}^{\sigma\,i}_{\lambda}[A({\mathbf z})]\,
\partial_i({\mathbf z})\,\right]\,({\Phi}^{-1}){}^{(\mu,y)}_
{\;\;\;\;(\sigma,z)}
({\mathbf y},{\mathbf z})=
{\delta}^{\mu}_{\lambda}\,{\delta}^3({\mathbf y}-{\mathbf z})\,.
\]
Notice that in the formula (\ref{5}), the matrix ${\Phi}^{-1}$,  
as well as the other terms of
the projector  $N$, is given on the gauge surface $ \Sigma$.


In our principal bundle, the 
 orbit metric $d_{(\mu,x)( \nu,y)}$ is determined by using the Killing vectors $K_{(\alpha , y)}$:
\[
d_{(\mu,x)( \nu,y)}=K^{(\alpha,i,z)}_{\;\;\;\;\;\;(\mu,x)}
G_{(\alpha,i,z)(\beta,j,u)}K^{(\beta,j,u)}_{\;\;\;\;\;\;( \nu,y)}+K^{(a,z)}_{\;\;\;\;(\mu, x)}G_{(a,z)(b,u)}K^{(b,u)}_{\;\;\;\;(\nu,y)} \]
That is,
\[
 d_{(\mu,x)( \nu,y)}=\Bigl[k_{\varphi \sigma}\delta^{kl}\tilde\mathcal D^{\varphi}_{\mu k}(A(\mathbf x))\mathcal D^{\sigma}_{\nu l}(A(\mathbf x))+
G_{ab}(\bar J_{\mu})^a_c (\bar J_{\nu})^b_{c'} f^c(\mathbf x)f^{c'}(\mathbf x)
\Bigr]\delta ^3(\mathbf x-\mathbf y)
\]
\[
 =\gamma_{\mu \nu}(\mathbf x,\mathbf y)+\gamma'_{\mu \nu}(\mathbf x,\mathbf y)
\]

An ``inverse matrix'' to the ``matrix'' $d_{(\mu,x)( \nu,y)}$ is defined by the following equation:
\[
 d_{(\mu,x)( \nu,y)}d^{(\nu,y)(\sigma,z)}=\delta^{( \sigma,z)}_{(\mu,x)}=
\delta ^{\sigma}_{\mu}\delta ^3(\mathbf z-\mathbf x).
\]
In explicit  form this equation is written as follows:
\begin{eqnarray*}
&&\Bigl[k_{\varphi \sigma}\delta^{kl}\tilde\mathcal D^{\varphi}_{\mu k}(A^{\ast}(\mathbf x))\mathcal D^{\sigma}_{\nu l}A^{\ast}((\mathbf x))+G_{ab}(\bar J_{\mu})^a_c (\bar J_{\nu})^b_{c'}{\tilde f}^c(\mathbf x){\tilde f}^{c'}(\mathbf x)
\Bigr]d^{(\nu,y)(\sigma,z)}
\nonumber\\
&&\;\;\;\;\;\;\;\;\;\;\;\;\;\;\;\;\;\;\;\;=\delta ^{\sigma}_{\mu}\delta ^3(\mathbf z-\mathbf x).
\end{eqnarray*}
Thus, $d^{(\nu,y)(\sigma,z)}$ is the Green function of the operator given by the expression in  square brackets. It is assumed  that a certain boundary condition for the equation is chosen.

The Green function $d^{(\nu,y)(\sigma,z)}$ and the Killing vectors are the  main elements with by which the ``Coulomb connection'' (or ``mechanical connection'') is determined:
$\hat{\omega}=\hat{\omega}^{\alpha}\otimes{\lambda}_{\alpha}$ in the principal fiber bundle ${\mathrm P}(\mathscr P\times_{\tilde\mathscr G}\mathscr V,\mathscr G)$:
\[
 \hat\omega ^{\alpha}=\bar{\rho}^{\alpha}_{\alpha '}(a(\mathbf x))\Bigl(\mathcal A^{(\alpha ',x)}_{\;\;\;\;(\beta,j, y)}\,d A^{\ast (\beta,j, y)}+\mathcal A^{(\alpha ',x)}_{\;\;\;(n, y)}d{\tilde f}^{(n,y)}\Bigr)+u^{\alpha}_{\mu}(a(\mathbf x))da^{\mu}(\mathbf x)),
\]
where the components of the connection are given by
\[
 \mathcal A^{(\alpha ',x)}_{\;\;\;\;(\beta,j, y)}=d^{(\alpha,x)(\sigma,z)}K^{(\mu, k, v)}_{\;\;\;\;\;\;(\sigma,z)}G_{(\mu,k,v)(\beta,j,y)}=k_{\mu \beta}[\mathcal D^{\mu}_{\sigma j}(A^{\ast}(\mathbf y))d^{(\alpha,x)(\sigma,z)}]
\]
and
\[
 A^{(\alpha ',x)}_{\;\;\;(p, y)}=d^{(\alpha,x)(\sigma,z)}K^{(a, v)}_{\;\;\;\;\;\;(\sigma,z)}G_{(a,v)(p,y)}=d^{(\alpha,x)(\sigma,z)}(\bar J_{\sigma})^a_c\tilde f^c(\mathbf y)G_{ap}.
\]

The following transformation of the coordinate basis in our principal bundle is connected with the replacement of the basis vector fields by the horizontal ones. This can be done with the help of horizontal projection operators, which are determined by the connection we have just defined. All this is similar to what we did in the finite-dimensional case. Therefore, we will not follow all the steps that ultimately must lead to the Lagrange-Poincaré equations in the functional space of gauge fields. Instead, for this purpose we will use the finite-dimensional equations (\ref{itog_hor_1}), (\ref{itog_hor_2}) and (\ref{itog_vert}). 

\section{The Lagrange-Poincaré equations in gauge theories}
The equations that we derive in this article are a special case of the Lagrange – Poincaré equations.
In this article, we restrict ourselves to a particular case of the Lagrange – Poincaré equations. They can be obtained from finite-dimensional equations if we assume that the expression under the projector $N^B_A$ in the first horizontal equation (\ref{itog_hor_1}) is zero. In addition, we neglect those terms of the first equations that explicitly depend on Killing vectors.
Then, from our assumption and the structure of the second horizontal equation (\ref{itog_hor_2}), it follows that the terms of the second equation with the projector $ N^r_B $ are equal to zero.
 Thus, we will deal with the following equations:
\begin{eqnarray}
&&\!\!\!\!\!\!\!\!\!\!\frac{d{\omega}^A}{dt}+G^{AR}\,{}^{\rm  H}\tilde {\Gamma}_{\tilde B\tilde MR}\,{\omega}^{\tilde B} {\omega}^{\tilde M} 
\nonumber\\
&&\qquad {} +G^{AR}\Bigl[{ \mathscr F}^{\alpha}_{\tilde Q\, R}  {\omega}^{\tilde Q} p_{\alpha} +\frac12({\mathscr D}_{ R} {d}^{\kappa\sigma}) p_{\kappa}p_{\sigma}+V_{,R}\Bigr]=0.                                 
\label{itog_hor_1_case}\\
&&{}\nonumber\\
&&\!\!\!\!\!\!\!\!\!\!\!\!\!\frac{d{\omega}^r}{dt}+G^{rm}\Bigl({}^{\rm  H}\tilde {\Gamma}_{\tilde A\tilde B m}{\omega}^{\tilde A} {\omega}^{\tilde B}
 +
{ \mathscr F}^{\alpha}_{\tilde Q m}  {\omega}^{\tilde Q} p_{\alpha}+\frac12({\mathscr D}_{m} {d}^{\kappa\sigma}) p_{\kappa}p_{\sigma}+V_{,m}\Bigr)=0.
\label{itog_hor_2_case}\\
&&{}\nonumber\\
&&\!\!\!\!\!\!\!\!\!\!\!\!\!\frac{d p_{\beta}}{dt}+ c^{\nu}_{\mu \beta}d^{\mu \sigma}p_{\sigma}p_{\nu}-c^{\nu}_{\sigma \beta}\mathscr A^{\sigma}_{\tilde E}\omega ^{\tilde E}p_{\nu}=0.
\label{itog_vert_case}
\end{eqnarray}
Since the Riemannian metric of the original manifold of gauge fields is flat, $ G_{AB} = \delta_{AB} $ must be used as a metric in these finite-dimensional equations. In addition, this fact must be taken into account when calculating the terms of equations are made with using the Killing relation.

In this regard, we first transform the terms of the equations so that later it was possible to replace them  by the appropriate functional expressions. Therms of equations with Christoffel symbols ${}^{\rm  H}\tilde {\Gamma}$, curvatures ${ \mathscr F}^{\alpha}$ and ${\mathscr D}_{m} {d}^{\kappa\sigma}$ will be expressed using the Killing vectors, the components of the mechanical connection and the metric on the orbit. Further we will list the obtained representations for these terms.

\begin{center}
\bf{Christoffel symbols for the  horizontal metric}
\end{center}
\begin{eqnarray*}
 G^{AR}{ }^{\;\;\rm H}{\Gamma}_{BMR}&=&-\frac12(\mathscr A^{\beta}_{B,M}K^A_{\beta}+\mathscr A^{\beta}_{M,B}K^A_{\beta})-(\mathscr A^{\beta}_{M}K^A_{\beta,B}+\mathscr A^{\beta}_{B}K^A_{\beta,M})\\
&{}& +\frac12(K^A_{\mu , D}K^D_{\sigma})(\mathscr A^{\mu}_{M}\mathscr A^{\sigma}_{B}+\mathscr A^{\sigma}_{M}\mathscr A^{\mu}_{B}).
{}\\
G^{AR}{ }^{\;\;\rm H}{\Gamma}_{BmR}&=&-\frac12\bigl(\mathscr A^{\beta}_{B,m}K^A_{\beta}+\mathscr A^{\beta}_{m,B}K^A_{\beta}\bigr)
-\mathscr A^{\beta}_{m}K^A_{\beta,B}\\
&{}& +\frac12(K^A_{\mu, D}K^D_{\sigma})(\mathscr A^{\mu}_{m}\mathscr A^{\sigma}_{B}+\mathscr A^{\sigma}_{m}\mathscr A^{\mu}_{B}).
{}\\
G^{AR}{ }^{\;\;\rm H}{\Gamma}_{pqR}&=&-\frac12\bigl(\mathscr A^{\beta}_{p,q}K^A_{\beta}+\mathscr A^{\beta}_{q,p}K^A_{\beta})\\
&{}& +\frac12(K^A_{\varepsilon, D}K^D_{\sigma})(\mathscr A^{\varepsilon}_{q}\mathscr A^{\sigma}_{p}+\mathscr A^{\sigma}_{q}\mathscr A^{\varepsilon}_{p}).
{}\\
G^{AR}{ }^{\;\;\rm H}{\Gamma}_{mBR}&=&-\frac12\bigl(\mathscr A^{\beta}_{m,B}K^A_{\beta}+\mathscr A^{\beta}_{B,m}K^A_{\beta}\bigr)
-\mathscr A^{\beta}_{m}K^A_{\beta,B}\\
&{}& +\frac12(K^A_{\mu, D}K^D_{\sigma})(\mathscr A^{\mu}_{m}\mathscr A^{\sigma}_{B}+\mathscr A^{\sigma}_{m}\mathscr A^{\mu}_{B}).
{}\\
G^{rm}{ }^{\;\;\rm H}{\Gamma}_{ABm}&=&-\frac12\bigl(\mathscr A^{\beta}_{A,B}K^r_{\beta}+\mathscr A^{\beta}_{B,A}K^r_{\beta})\\
&{}& +\frac12(K^r_{\mu, p}K^p_{\sigma})(\mathscr A^{\sigma}_{A}\mathscr A^{\mu}_{B}+\mathscr A^{\mu}_{A}\mathscr A^{\sigma}_{B}).
{}\\
G^{rm}{ }^{\;\;\rm H}{\Gamma}_{pBm}&=&-\frac12\bigl(\mathscr A^{\beta}_{p,B}K^r_{\beta}+\mathscr A^{\beta}_{B,p}K^r_{\beta})-\mathscr A^{\beta}_{B}K^r_{\beta,p}\\
&{}& +\frac12(K^r_{\varepsilon, q}K^q_{\mu})(\mathscr A^{\mu}_{p}\mathscr A^{\varepsilon}_{B}+\mathscr A^{\mu}_{B}\mathscr A^{\varepsilon}_{p}).
{}\\
G^{rm}{ }^{\;\;\rm H}{\Gamma}_{pqm}&=&-\frac12\bigl(\mathscr A^{\beta}_{p,q}K^r_{\beta}+\mathscr A^{\beta}_{q,p}K^r_{\beta})
-(\mathscr A^{\beta}_{p}K^r_{\beta,q}+\mathscr A^{\beta}_{q}K^r_{\beta,p})\\
&{}& +\frac12(K^r_{\mu, n}K^n_{\nu})(\mathscr A^{\mu}_{q}\mathscr A^{\nu}_{p}+\mathscr A^{\nu}_{q}\mathscr A^{\mu}_{p}).
\end{eqnarray*}

\begin{center}
\bf{Curvatures $\mathbf{G^{\tilde A\tilde E}\mathscr F^{\alpha}_{\tilde Q \tilde E}}$}
\end{center}
\begin{eqnarray*}
 &&G^{AE}{\mathscr F}^{\alpha}_{QE}=-(K^S_{\varphi,Q})(d^{\varphi \alpha}\mathscr A^{\mu}_S+d^{\varphi \mu}\mathscr A^{\alpha}_S)K^A_{\mu}-(K^A_{\epsilon ,B }K^B_{\nu})({d}^{\alpha \epsilon}{\mathscr A}^{\nu}_{Q}+{d}^{\alpha \nu }{\mathscr A}^{\epsilon}_{Q})
\nonumber\\
&&+2{d}^{\,\alpha \mu}K^A_{\mu, Q}+c^{\alpha}_{\nu\mu}{d}^{\mu\varphi}{\mathscr A}^{\nu}_{Q}K^A_{\varphi}.
\end{eqnarray*}

\begin{eqnarray*}
G^{AR}{\mathscr F}^{\alpha}_{qR}&=&-(K^r_{\mu,q})(d^{\mu \alpha}\mathscr A^{\varphi}_r+d^{\mu\varphi}\mathscr A^{\alpha}_r)K^A_{\varphi}-(K^A_{\nu ,B }K^B_{\varphi})({d}^{\alpha \nu}{\mathscr A}^{\varphi}_{q}+{d}^{\alpha \varphi }{\mathscr A}^{\nu}_{q})
\nonumber\\
&{ }&+c^{\alpha}_{\nu\mu}{d}^{\mu\varphi}{\mathscr A}^{\nu}_{q}K^A_{\varphi}.
\end{eqnarray*}

\begin{eqnarray*}
 G^{rm}{\mathscr F}^{\alpha}_{Q\,m}&=&-K^T_{\mu, Q}({d}^{\alpha \mu}{\mathscr A}^{\beta}_{T}+{d}^{\beta \mu}{\mathscr A}^{\alpha}_{T})K^r_{\beta}
-(K^n_{\nu }K^r_{\mu,n})\,({d}^{\alpha \mu}{\mathscr A}^{\nu}_{Q}+{d}^{\alpha \nu}{\mathscr A}^{\mu}_{Q})
\nonumber\\
&{}&+c^{\alpha}_{\nu \mu}\,{d}^{\mu \beta}{\mathscr A}^{\nu}_{Q}\,K^r_{\beta}.
\end{eqnarray*}

\begin{eqnarray*}
 G^{rm}{\mathscr F}^{\alpha}_{q\,m}&=&-K^n_{\mu, q}({d}^{\alpha \mu}{\mathscr A}^{\beta}_{n}+{d}^{\beta \mu}{\mathscr A}^{\alpha}_{n})K^r_{\beta}
-(K^p_{\nu }K^r_{\mu,p})\,({d}^{\,\alpha \mu}{\mathscr A}^{\nu}_{q}+{d}^{\,\alpha \nu}{\mathscr A}^{\mu}_{q})
\nonumber\\
&{}&+2{d}^{\,\alpha \beta} K^r_{\beta, q}   +c^{\alpha}_{\nu \mu}\,{d}^{\mu \beta}{\mathscr A}^{\nu}_{q}\,K^r_{\beta}.
\end{eqnarray*}

\begin{center}
{$\mathbf {G^{\tilde A\tilde R}({\mathscr D}_{\tilde R}d^{\kappa\sigma})p_{\kappa}p_{\sigma}}$}
\end{center}
\[
 G^{AR}({\mathscr D}_Rd^{\kappa\sigma})p_{\kappa}p_{\sigma}=2\,\Bigl[(K^D_{\beta}K^A_{\mu,D})\,d^{\beta\kappa}d^{\mu\sigma}+c^{\kappa}_{\beta \mu}d^{\beta\epsilon}d^{\mu\sigma}K^A_{\epsilon}\Bigr]p_{\kappa}p_{\sigma}
\]

\[
 G^{rm}({\mathscr D}_md^{\kappa\sigma})p_{\kappa}p_{\sigma}=2\,\Bigl[(K^n_{\beta}K^r_{\mu,n})\,d^{\beta\kappa}d^{\mu\sigma}+c^{\kappa}_{\beta \mu}d^{\beta\epsilon}d^{\mu\sigma}K^r_{\epsilon}\Bigr]p_{\kappa}p_{\sigma}
\]

Note that before starting the transition to the functional representation in the Christoffel symbols, the partial derivatives of the connections are replaced using the following formulas:
\begin{center}
\bf{ Partial derivatives $\mathbf{\mathscr A^{\alpha}_{\tilde Q,\tilde R}}$}
\end{center}
\begin{eqnarray*}
&& \mathscr A^{\alpha}_{Q,R}=-{d}^{\alpha \epsilon}\,K^A_{\epsilon\, R}\,G_{AB}\,K^B_{\mu}{\mathscr A}^{\mu}_{Q}-{\mathscr A}^{\alpha}_{B}\,{\mathscr A}^{\mu}_{Q}\,K^B_{\mu\, R}+{d}^{\alpha \mu}\,K^B_{\mu\, R}\,G_{BQ}\\
&& { }\\
&& \mathscr A^{\alpha}_{Q,p}=-{d}^{\alpha \varepsilon}\,K^r_{\varepsilon \,p}\,G_{rn}\,K^n_{\mu}{\mathscr A}^{\mu}_{Q}-{\mathscr A}^{\alpha}_{n}\,{\mathscr A}^{\mu}_{Q}\,K^n_{\mu\, p}\\
&&{ }\\
&&\mathscr A^{\alpha}_{p,Q}=
-{d}^{\alpha \varepsilon}\,K^A_{\varepsilon\, Q}\,G_{AB}\,K^B_{\mu}{\mathscr A}^{\mu}_{p}-{\mathscr A}^{\alpha}_{B}\,{\mathscr A}^{\mu}_{p}\,K^B_{\mu\, Q}\\
&&{ }\\
&&\mathscr A^{\alpha}_{p,q}=-{d}^{\alpha \varepsilon}\,K^r_{\varepsilon\, q}\,G_{rn}\,K^n_{\mu}{\mathscr A}^{\mu}_{p}-{\mathscr A}^{\alpha}_{n}\,{\mathscr A}^{\mu}_{p}\,K^n_{\mu\, q}+{d}^{\alpha \mu}\,K^m_{\mu\, q}\,G_{mp}
\end{eqnarray*}

Another equivalent representation of derivatives are

\[
\mathscr A^{\beta}_{B,m}=2{d}^{\beta \mu}(K^q_{\mu}K^p_{\varphi,q})G_{pm}\mathscr A^{\varphi}_B+c^{\sigma}_{\varphi \mu}{d}^{\beta \mu}K^p_{\sigma}\mathscr A^{\varphi}_BG_{pm},
\]

\[
\mathscr A^{\beta}_{m,B}=2{d}^{\beta \mu}(K^E_{\mu}K^D_{\varphi,E})G_{BD}\mathscr A^{\varphi}_m+c^{\sigma}_{\varphi \mu}{d}^{\beta \mu}K^D_{\sigma}\mathscr A^{\varphi}_mG_{BD}.
\]

To obtain a functional representation for the members of the equations, one needs to treat the indices of variables as if they were compact notations of extended indices.
\[
 A\to(\alpha, i,x);\quad a\to(n,y);\quad \mu\to (\mu.u);\ldots \rm{etc.}
\]
Recall that our basic variables are $\omega ^A\equiv \dot Q^{\ast}{}^A$, $\omega ^n\equiv \dot {\tilde f^n}$. So, we have
\[
 \omega ^A(t)\to\frac{d}{dt}A^{\ast(\alpha,i,x)}(t)\equiv \frac{d}{dt}A^{\ast \alpha,i}(\mathbf x,t)\equiv\dot A^{\ast \alpha i}(\mathbf x,t),
\]
and similarly for $\omega ^n(t)$: $\omega ^n(t)\to\frac{d}{dt}\tilde f^{(n,x)}(t)\equiv\dot {\tilde f^n}(\mathbf x,t)$.

To obtain functional representations for the terms of the equations, we made the following replacements:

$$K^A_{\beta, B}\to K^{(\alpha , i, x)}_{\;\;\;\;(\beta,u)(\epsilon, k,z)},$$
where

\[
 K^{(\alpha , i, x)}_{\;\;\;\;(\beta,u)(\epsilon, k,z)}\equiv
\frac{\delta}{\delta A^{(\epsilon,k,z)}}\,
K^{(\alpha , i,x)}_{\;\;\;\;(\beta,u)} =\delta^i_k c^{\alpha}_{\epsilon\beta}\delta^3(\mathbf x-\mathbf u)\delta^3(\mathbf x-\mathbf z).
\]

$K^A_{\mu, E}K^E_{\nu}\to K^{(\alpha,i,x)}_{(\beta,v)(...)}K^{(...)}_{(\nu,u)}$
\[
 K^{(\alpha,i,x)}_{(\beta,v)(...)}K^{(...)}_{(\nu,u)}=c^{\alpha}_{\mu' \beta}\bigl[\mathcal D^{\mu' i}_{\nu}(A^{\ast}(x))\delta^3(\mathbf x-\mathbf u)\bigr]\delta^3(\mathbf x-\mathbf v)
\]

$K^n_{\beta, m}\to K^{(n,v)}_{\;\;(\beta,y)(m,z)}$

\[
 K^{(n,v)}_{\;\;\;(\beta,y)(m,z)}={(\bar J_{\beta}})^n_m\,\delta^3(\mathbf v-\mathbf z)\delta^3(\mathbf v-\mathbf y)
\]

$K^n_{\beta,m}K^{m}_{\alpha}\to K^{(n,v)}_{\;\;(\beta,y)(m,z)}K^{(m,z)}_{\;\;\;(\alpha, u)}$

\[
 K^{(n,v)}_{\;\;(\beta,y)(m,z)}K^{(m,z)}_{\;\;\;(\alpha, u)}=(\bar J_{\beta})^n_m(\bar J_{\alpha})^m_qf^q(\mathbf y)\delta^3(\mathbf v-\mathbf y)\delta^3(\mathbf v-\mathbf u)
\]

To get expressions in the right parts of the formulas, you need to take
sum over repeated generalized indices.  Sum over continuous indices means corresponding integration.

There are also the appropriate functional representations for the connections.

$\mathscr A^{\alpha}_B\to\mathscr A^{(\alpha, x)}_{\;\;\;\;(\beta,j,y)}$

\[
 \mathscr A^{(\alpha, x)}_{\;\;\;\;(\beta,j,y)}=[\mathcal D^{\varphi}_{\mu j}(A^{\ast}(y))d^{(\alpha, x)(\mu,y)}]k_{\varphi \beta}
\]

$\mathscr A^{\alpha}_m\to\mathscr A^{(\alpha, x)}_{\;\;\;(m,z)}$

\[
 \mathscr A^{(\alpha, x)}_{\;\;\;(m,z)}=d^{(\alpha, x)(\beta'\!, z)}({\bar J_{\beta'}})^n_pf^p(\mathbf x)G_{nm}
\]
But in our final formulas we do not use them, despite the fact that this can be done as it does not lead to the simplification of the already rather complex expressions. 

The results of our calculations - functional representations of the terms of the equations are presented in Appendix.

The first horizontal equation  (\ref{itog_hor_1_case}) results in to the following Lagrange-Poincar\'{e} equation for the gauge system
\begin{eqnarray}
 &&\!\!\!\!\!\!\!\!\!\!\frac{d{\omega}^{(\alpha, i,x)}}{dt}+\underline{G^{AR}\,{}^{\rm  H}\tilde {\Gamma}_{ B MR}\,{\omega}^{ B} {\omega}^{ M} :} \nonumber\\
&&-2\,c^{\alpha}_{\varepsilon \beta}\Bigl[\;\int\!d^3\!y\,\mathscr A^{(\beta,x)}_{\;\;\;\;(\gamma,j,\,y)}\,\omega^{(\gamma,j,y)}\Bigr]\omega^{(\varepsilon,i,x)}
\nonumber\\
&& +c^{\alpha}_{\varphi \mu}\!\int\!d^3\!yd^3\!z\,\mathscr A^{(\mu,x)}_{\;\;\;\;(\gamma,j,\,y)}\Bigl[\mathcal D^{\varphi i}_{\nu}(A^{\ast}(\mathbf x))\mathscr A^{(\nu,x)}_{\;\;\;\;(\varepsilon,k,\,z)}\Bigr]\omega^{(\gamma,j,y)}\omega^{(\varepsilon,k,z)}
\nonumber\\
&&+\underline{2G^{AR}\,{}^{\rm  H}\tilde {\Gamma}_{B m R}\,{\omega}^{ B} {\omega}^{ m}{}:}  \nonumber\\
&&-2\delta^i_k c^{\alpha}_{\epsilon \beta}\biggl(\int\! d^3v\,\mathscr A^{(\beta,x)}_{\;\;\;(m,v)}\omega^{(m,v)}\biggr)\omega^{(\epsilon,k,x)}\nonumber\\
&&+c^{\alpha}_{\epsilon \beta}\int\!d^3v\,d^3z\biggl[\mathscr A^{(\beta,x)}_{\;\;\;\;(\epsilon,k,z)}\overleftarrow{\overrightarrow{\mathcal D}}{}^{\mu i}_{\nu}(A^{\ast}(\mathbf x))\mathscr A^{(\nu,x)}_{\;\;\;(m,v)}\omega^{(m,v)}
\omega^{(\epsilon,k,z)}\biggr]
\nonumber\\
&&+\underline{G^{AR}\,{}^{\rm  H}\tilde {\Gamma}_{p q R}\,{\omega}^{ p} {\omega}^{ q}{}:}
\nonumber\\
&&\frac12c^{\alpha}_{\mu \beta}\!\int\! d^3\!yd^3\!z\Bigl(A^{(\nu,x)}_{\;\;(p,y)}\overleftarrow{\overrightarrow{\mathcal D}}{}^{\,\mu i}_{\nu}(A^{\ast}(\mathbf x))\mathscr A^{(\beta,x)}_{\;\;(q,z)}\Bigr) \omega^{(p,y)}\omega^{(q,z)}
\nonumber\\
&&+\underline{G^{AR}{ \mathscr F}^{\alpha}_{Q\, R}  {\omega}^{Q} p_{\alpha}:}
\nonumber\\
&& -c^{\alpha}_{\mu'\varphi}\!\int\!d^3\!u\,d^3\!z\,d^{(\alpha' ,u)(\varphi,x)}\Bigl[\mathcal D^{\mu' i}_{\nu}(A^{\ast}(x))
\mathscr A^{(\nu,x)}_{\;\;\;\;(\varepsilon,k,z)}\Bigr]\omega^{(\varepsilon,k,z)}\,p_{(\alpha',u)}\nonumber\\
&& -c^{\alpha}_{\mu'\varphi}\!\int\!d^3\!u\,d^3\!z\,\mathscr A^{(\varphi,x)}_{\;\;\;\;(\varepsilon,k,z)}
\Bigl[\mathcal D^{\mu' i}_{\nu}(A^{\ast}(x))d^{(\alpha' ,u)(\nu,x)}\Bigr]\omega^{(\varepsilon,k,z)}\,p_{(\alpha',u)}
\nonumber\\
&&+ 2c^{\alpha}_{\varepsilon \mu}\Bigl[\;\;\int\!d^3\!u\,d^{(\beta ,u)(\mu,x)}p_{(\beta,u)}\Bigr]\omega^{(\varepsilon,i,x)}
\nonumber\\
&&+ c^{\alpha'}_{\nu\mu}\!\int\!d^3\!u\,d^3\!z\,\mathscr A^{(\nu,u)}_{\;\;\;\;(\varepsilon,k,z)}\Bigl[\mathcal D^{\alpha i}_{\varphi}(A^{\ast}(x))d^{(\mu ,u)(\varphi,x)}\Bigr]\omega^{(\varepsilon,k,z)}\,p_{(\alpha',u)}
\nonumber\\
&&+\underline{G^{AR}{ \mathscr F}^{\alpha}_{q\, R}  {\omega}^{q} p_{\alpha}:}
\nonumber\\
&&-c^{\alpha}_{\mu' \nu}\!\int\!d^3\!u\,d^3\!y\,d^{(\alpha',u)(\nu,x)}\Bigl[\mathcal D^{\mu' i}_{\varphi}(A^{\ast}(x))\mathscr A^{(\varphi,x)}_{\;\;\;\;(q,y)}\Bigr]\,\omega^{(q,y)}p_{(\alpha',u)}
\nonumber\\
&&-c^{\alpha}_{\mu' \nu}\!\int\!d^3\!u\,d^3\!y\,\mathscr A^{(\nu,x)}_{\;\;\;\;(q,y)}\Bigl[\mathcal D^{\mu' i}_{\varphi}(A^{\ast}(x))d^{(\alpha',u)(\varphi,x)}\Bigr]\,\omega^{(q,y)}p_{(\alpha',u)}
\nonumber\\
&&+c^{\alpha'}_{\nu \mu}\!\int\!d^3\!u\,d^3\!y\,\mathscr A^{(\nu,u)}_{\;\;\;\;(q,y)}\Bigl[\mathcal D^{\alpha i}_{\varphi}(A^{\ast}(x))d^{(\mu,u)(\varphi,x)}\Bigr]\,\omega^{(q,y)}p_{(\alpha',u)}
\nonumber\\
&&+\underline{\frac12({\mathscr D}_{ R} {d}^{\kappa\sigma}) p_{\kappa}p_{\sigma}:}
\nonumber\\
&&+c^{\alpha}_{\mu' \mu}\!\int\!d^3\!z\,d^3\!z'\,d^{(\mu,x)(\sigma,z')}\Bigl[\mathcal D^{\mu' i}_{\beta}(A^{\ast}(x))d^{(\beta,x)(\kappa,z)}\Bigr]p_{(\kappa,z)}p_{(\sigma,z')}
\nonumber\\
&&+\underline{G^{AR}V_{,R}:}
\nonumber\\
&&
-\mathcal D^{\alpha}_{\beta j}(A^{\ast}(\mathbf x,t))F^{\beta j i}(\mathbf x,t)+G_{ab}k^{\alpha \gamma}(\bar J_{\gamma})^a_d \,\tilde f^d(\mathbf x,t)({\nabla}\tilde f)^{b i}(\mathbf x,t)=0
\label{itog_hor_1_gauge}
\end{eqnarray}

The second horizontal Lagrange-Poincar\'{e} equation for the gauge system which follows from (\ref{itog_hor_2_case}) is
\begin{eqnarray*}
&&\frac{d{\omega}^{(r,x)}(t)}{dt}+\underline{G^{rm}{}^{\rm  H}\tilde {\Gamma}_{ A B m}{\omega}^{A} {\omega}^{B}:}\nonumber\\
&&+ \Bigl(c^{\nu}_{\gamma \mu} (\bar J_{\beta})^r_{p}\tilde f^p(\mathbf x)\,k_{\sigma\alpha}\!\int\!d^3\!zd^3\!z'
d^{(\beta,x)(\mu,z)}\,\Bigl[{\mathcal D}^{\sigma }_{\epsilon n}(A^{\ast}(\mathbf z'))\mathscr A^{(\epsilon,z')}_{\;\;\;\;(\nu,k,z)}\Bigr]\Bigr.
\nonumber\\
&&
 +c^{\nu}_{\gamma \mu}k_{\varphi \nu} (\bar J_{\beta})^r_{p}\tilde f^p(\mathbf x)\!\int\! d^3\!zd^3\!z'\mathscr A^{(\mu,z)}_{\;\;\;\;(\alpha,n,z')}\mathscr A^{(\beta,x)}_{\;\;\;\;(\nu,k,z)}
\nonumber\\
&&
  -c^{\sigma}_{\gamma \epsilon}\delta_{kn}k_{\sigma \alpha}(\bar J_{\beta})^r_{p}\,\tilde f^p(\mathbf x)\!\int \!d^3\!z\,d^{(\beta,x)(\epsilon, z)}
\nonumber\\
 &&
  \Bigl.+(\bar J_{\mu})^r_{m'}(\bar J_{\epsilon})^{m'}_q \tilde f^q(\mathbf x)\!\int\! d^3\!zd^3\!z'\mathscr A^{(\epsilon,x)}_{\;\;\;\;(\alpha,n,z')}
\mathscr A^{(\mu,x)}_{\;\;\;\;(\gamma,k,z)}\Bigr)
\omega^{(\alpha,n,z')}\omega^{(\gamma,k,z)}
 \nonumber\\
&&+2\underline{G^{rm}{}^{\rm  H}\tilde {\Gamma}_{ p B m}{\omega}^{p} {\omega}^{B}:}\nonumber\\
&&+ \frac12(\bar J_{\varepsilon'})^{q}_{p}G_{q,n}(\bar J_{\mu})^{n}_{n'}\tilde f^{n'}\!(\mathbf x)(\bar J_{\beta})^{r}_{l}\tilde f^l(\mathbf x)\!\int\! d^3\!zd^3\!z'd^{(\beta,x)(\varepsilon',z')}\mathscr A^{(\mu,z')}_{\;\;\;(\varepsilon,k,z)}\omega^{(\varepsilon,k,z)}\omega^{(p,z')}\nonumber\\
&&+\frac12(\bar J_{\mu})^{n}_{p} (\bar J_{\beta})^{r}_{q}\tilde f^q(\mathbf x)\!\int\! d^3\!zd^3\!z'  \mathscr A^{(\beta,x)}_{\;\;(n,z')}            \mathscr A^{(\mu,z')}_{\;\;\;(\varepsilon,k,z)}\omega^{(\varepsilon,k,z)}\omega^{(p,z')}
\nonumber\\
&& +\frac12c^{\gamma}_{\varepsilon\varepsilon'}k_{\gamma \nu}(\bar J_{\beta})^{r}_{q}\tilde f^q(\mathbf x)\delta_{kk'}\!\int\! d^3\!zd^3\!z'd^{(\beta,x)(\varepsilon',z)}\Bigl[\mathcal D^{\nu k'}_{\mu}(A^{\ast}(z))\mathscr A^{(\mu,z)}_{\;\;(p,z')}\Bigr]\omega^{(\varepsilon,k,z)}\omega^{(p,z')}
\nonumber\\
&& +\frac12c^{\gamma}_{\varepsilon \mu}(\bar J_{\beta})^{r}_{q}\tilde f^q(\mathbf x)\!\int\! d^3\!zd^3\!z'\mathscr A^{(\beta,x)}_{\;\;\;(\gamma,k,z)}\mathscr A^{(\mu,z)}_{\;\;\;(p,z')}\,\omega^{(\varepsilon,k,z)}\omega^{(p,z')}\\
\nonumber\\
&&-(\bar J_{\beta})^{r}_{p}\Bigl[\int \!d^3\!z \,\mathscr A^{(\beta,x)}_{\,\;\;\;(\varepsilon,k,z)}\,\omega^{(\varepsilon,k,z)}\Bigr]  \omega^{(p,x)}           
\nonumber\\
&&
+\frac12(\bar J_{\gamma})^{r}_{k}(\bar J_{\mu})^{k}_{n}\tilde f^n(\mathbf x)\!\int\! d^3\!zd^3\!z'\Bigl( \mathscr A^{(\mu,x)}_{\;\;\;(p,z')}\mathscr A^{(\gamma,x)}_{\,\;\;\;(\varepsilon,k,z)}+\mathscr A^{(\gamma,x)}_{\;\;\;(p,z')}\mathscr A^{(\mu,x)}_{\,\;\;\;(\varepsilon,k,z)}\Bigr)\omega^{(\varepsilon,k,z)}\omega^{(p,z')}
\nonumber\\
&&+\underline{G^{rm}{}^{\rm  H}\tilde {\Gamma}_{ p q m}{\omega}^{p} {\omega}^{q}:}\nonumber\\
&& +(\bar J_{\beta})^{r}_{n}\tilde f^{n}(\mathbf x)(\bar J_{\varepsilon})^{r'}_{q}G_{r'n'}(\bar J_{\mu})^{n'}_{m'}\!\int\! d^3\!zd^3\!y\,d^{(\beta,x)(\varepsilon,z)}\tilde f^{m'}(\mathbf z)\,\mathscr A^{(\mu,z)}_{\;\;\;(p,y)}\,\omega^{(p,y)}\omega^{(q,z)}
\nonumber\\
&&+ (\bar J_{\mu})^{n}_{q}(\bar J_{\beta})^{r}_{n'}\tilde f^{n'}(\mathbf x)\!\int\!d^3\!zd^3\!y\,\mathscr A^{(\beta,x)}_{\;\;\;(n,z)}\mathscr A^{(\mu,z)}_{\;\;\;(p,y)}\,\omega^{(p,y)}\omega^{(q,z)}
\nonumber\\
&&
 -(\bar J_{\mu})^{n}_{q}G_{np}(\bar J_{\beta})^{r}_{n'}\tilde f^{n'}(\mathbf x)\!\int\!d^3\!y\, d^{(\beta,x)(\mu,y)}\,\omega^{(p,y)}\omega^{(q,y)}
\nonumber\\
&&-2(\bar J_{\beta})^{r}_{q}\biggl[\;\int\!d^3\!y\,\mathscr A^{(\beta,z)}_{\;\;\;(p,y)}\,\omega^{(p,y)}\biggr]\omega^{(q,x)}
\nonumber\\
&&
+ (\bar J_{\mu})^{r}_{m}(\bar J_{\nu})^{m}_{k}\tilde f^{k}(\mathbf x)\!\int\!d^3\!z\,d^3\!y\,\mathscr A^{(\mu,x)}_{\;\;\;(q,z)}\mathscr A^{(\nu,x)}_{\;\;\;(p,y)}\,\omega^{(p,y)}\omega^{(q,z)}
\nonumber\\
&&\underline{{ \mathscr F}^{\alpha}_{\tilde Q m}  {\omega}^{\tilde Q} p_{\alpha}:}\nonumber\\
&&
-c^{\gamma}_{\varepsilon \mu}(\bar J_{\beta})^{r}_{m}\tilde f^m(\mathbf x)\!\int\!d^3\!u\,d^3\!z\,d^{(\alpha',u)(\mu,z)}\,\mathscr A^{(\beta,x)}_{\;\;\;\;(\gamma,k,z)}\,\omega^{(\varepsilon,k,z)}p_{(\alpha',u)}
\nonumber\\
&&
-c^{\gamma}_{\varepsilon \mu}(\bar J_{\beta})^{r}_{m}\tilde f^m(\mathbf x)\!\int\!d^3\!u\,d^3\!z\,d^{(\beta,x)(\mu,z)}\,\mathscr A^{(\alpha',u)}_{\;\;\;\;(\gamma,k,z)}\,\omega^{(\varepsilon,k,z)}p_{(\alpha',u)}
\nonumber\\
&&-(\bar J_{\mu})^{r}_{m'}(\bar J_{\nu})^{m'}_{q}\tilde f^q(\mathbf x)\!\int\!d^3\!u\,d^3\!z\,d^{(\alpha',u)(\mu,x)}\,\mathscr A^{(\nu,x)}_{\;\;\;\;(\varepsilon,k,z)}\,\omega^{(\varepsilon,k,z)}p_{(\alpha',u)}
\nonumber\\
&&
-(\bar J_{\mu})^{r}_{m'}(\bar J_{\nu})^{m'}_{q}\tilde f^q(\mathbf x)\!\int\!d^3\!u\,d^3\!z\,d^{(\alpha',u)(\nu,x)}\,\mathscr A^{(\mu,x)}_{\;\;\;\;(\varepsilon,k,z)}\,\omega^{(\varepsilon,k,z)}p_{(\alpha',u)}
\nonumber\\
&&
+c^{\alpha'}_{\nu \mu}(\bar J_{\beta})^{r}_{n}\tilde f^n(\mathbf x)\!\int\!d^3\!u\,d^3\!z\,d^{(\mu,u)(\beta,x)}\,\mathscr A^{(\nu,u)}_{\;\;\;\;(\varepsilon,k,z)}\,\omega^{(\varepsilon,k,z)}p_{(\alpha',u)}
\nonumber\\
&&\underline{G^{rm}{\mathscr F}^{\alpha'}_{q\,m}\,\omega^q\,p_{\alpha'}:}\nonumber\\
&&
-(\bar J_{\mu})^{n}_{q}(\bar J_{\beta})^{r}_{m}\tilde f^m(\mathbf x)\!\int\!d^3\!u\,d^3\!z\,\mathscr A^{(\beta,x)}_{\;\;\;(n,z)}d^{(\alpha',u)(\mu,z)}\,\,\omega^{(q,z)}p_{(\alpha',u)}
\nonumber\\
&&
-(\bar J_{\mu})^{n}_{q}(\bar J_{\beta})^{r}_{m}\tilde f^m(\mathbf x)\!\int\!d^3\!u\,d^3\!z\,d^{(\beta,x)(\mu,z)}\mathscr A^{(\alpha',u)}_{\;\;\;\;(n,z)}\,\,\omega^{(q,z)}p_{(\alpha',u)}
\nonumber\\
&&
-(\bar J_{\mu})^{r}_{m'}(\bar J_{\nu})^{m'}_{n'}\tilde f^{n'}(\mathbf x)\!\int\!d^3\!u\,d^3\!z\,d^{(\alpha',u)(\mu,x)}\mathscr A^{(\nu,x)}_{\;\;\;\;(q,z)}\,\,\omega^{(q,z)}p_{(\alpha',u)}
\nonumber\\
&&
-(\bar J_{\mu})^{r}_{m'}(\bar J_{\nu})^{m'}_{n'}\tilde f^{n'}(\mathbf x)\!\int\!d^3\!u\,d^3\!z\,d^{(\alpha',u)(\nu,x)}\mathscr A^{(\mu,x)}_{\;\;\;\;(q,z)}\,\,\omega^{(q,z)}p_{(\alpha',u)}
\nonumber\\
&&+2\,(\bar J_{\beta})^{r}_{q}\,\omega^{(q,x)}\!\int\!d^3\!u\,\,d^{(\alpha',u)(\beta,x)}\,p_{(\alpha',u)}
\nonumber\\
&&+c^{\alpha'}_{\nu \mu}\,(\bar J_{\beta})^{r}_{m}\tilde f^{m}(\mathbf x)\!\int\!d^3\!u\,d^3\!z\,d^{(\mu,u)(\beta,x)}\mathscr A^{(\nu,x)}_{\;\;\;\;(q,z)}\,\,\omega^{(q,z)}p_{(\alpha',u)}
\nonumber\\
&&\underline{\frac12 G^{rm}({\mathscr D}_md^{\kappa\sigma})p_{\kappa}p_{\sigma}:}\nonumber\\
&&
+\,(\bar J_{\mu})^{r}_{n}(\bar J_{\beta})^{n}_{q}\tilde f^{q}(\mathbf x)\!\int\!d^3\!z\,d^3\!z'\,d^{(\beta,x)(\kappa,z)}d^{(\mu,x)(\sigma,z')}p_{(\kappa,z)}p_{(\sigma,z')}
\nonumber\\
&&+\,c^{\kappa}_{\beta \mu}\,(\bar J_{\varepsilon})^{r}_{n}\tilde f^{n}(\mathbf x)\!\int\!d^3\!z\,d^3\!z'\,d^{(\beta,z)(\varepsilon,x)}d^{(\mu,z)(\sigma,z')}p_{(\kappa,z)}p_{(\sigma,z')}
\nonumber\\
&&\underline{G^{rm}V_{,m}:} +G^{rm}G_{ab}\tilde{\nabla}^{a i }_m(\nabla_i \tilde f^b)=0.
\label{itog_hor_2_gauge}\\
\end{eqnarray*}

The vertical Lagrange-Poincar\'{e} equation for the gauge system is  
\begin{eqnarray*}
&&\frac{d p_{\beta}(\mathbf x,t)}{dt}+ c^{\nu}_{\mu \beta}\Bigl(\!\int\!d^3\!y\,d^{(\mu,x) (\sigma,y)}p_{\sigma}(\mathbf y,t)\Bigr)p_{\nu}(\mathbf x,t)\nonumber\\
&&-c^{\nu}_{\sigma \beta}\Bigl(\!\int\!d^3\!z\mathscr A^{(\sigma,x)}_{\;\;\;(\varepsilon,k,z)}\dot A^{\ast}{}^{\varepsilon k)}\Bigr) p_{\nu}(\mathbf x,t)-c^{\nu}_{\sigma \beta}\Bigl(\!\int\!d^3\!y\mathscr A^{(\sigma,x)}_{\;\;(n,y)}\dot{\tilde f^n}(\mathbf y,t)\Bigr)p_{\nu}(\mathbf x,t)=0
\end{eqnarray*}

\section{Concluding remarks}
The obtained equations are represented a rather complex expressions. A possible simplification of these equations can be obtained by projection of the equations onto the orbit space of the principal bundle.This is achieved by setting the group variable to zero.
This suggests that the full equations apparently describe the dynamics of the system in the excited state.A direct consequence of these equations are the equations for the relative equilibria of the dynamical system under consideration. These equations can easily be derived from the result of those equations that are obtained in the work. It should also be noted that the role played by the equations of the mechanical system. Without these equations, it would be difficult to understand in complex expression the nature and origin of individual terms of the equation for the gauge system. It remains unclear whether it is possible to somehow simplify the resulting equation based on some kind of symmetry. It is also not clear whether the equation for relative equilibria can be used to study the problem related to  symmetry breaking.

\appendix
\section*{Appendix A}
\section*{Functional representation for  terms of the Lagrange-Poincar\'{e} equations}
\setcounter{equation}{0}
\def\theequation{A.\arabic{equation}}

\begin{eqnarray*}
 \mathbf{G^{AR}{ }^{\;\;\rm H}{\Gamma}_{BMR}}&=&-\frac12(\mathscr A^{\beta}_{B,M}K^A_{\beta}+\mathscr A^{\beta}_{M,B}K^A_{\beta})-(\mathscr A^{\beta}_{M}K^A_{\beta,B}+\mathscr A^{\beta}_{B}K^A_{\beta,M})\\
&{}& +\frac12(K^A_{\mu , D}K^D_{\sigma})(\mathscr A^{\mu}_{M}\mathscr A^{\sigma}_{B}+\mathscr A^{\sigma}_{M}\mathscr A^{\mu}_{B})
\end{eqnarray*}

$\mathscr A^{\beta}_{B,M}=-{d}^{\beta \epsilon}\,K^{A'}_{\epsilon\,, M}\,G_{A'B'}\,K^{B'}_{\mu}{\mathcal A}^{\mu}_{B}-{\mathcal A}^{\beta}_{D}\,{\mathcal A}^{\mu}_{B}\,K^D_{\mu\,, M}+{d}^{\beta \mu}\,K^D_{\mu\,, M}\,G_{DB}$

The expression ${\Roman 1}=-\frac12(\mathscr A^{\beta}_{B,M}K^A_{\beta}+\mathscr A^{\beta}_{M,B}K^A_{\beta})K^A_{\beta}\,\omega^B\omega^M=-\mathscr A^{\beta}_{B,M}K^A_{\beta}K^A_{\beta}\,\omega^B\omega^M$ consists of three terms:\\
$\mathbf{\Roman 1_{1)}}:$
\begin{eqnarray*}
&&c^{\nu}_{\gamma \epsilon'}k_{\nu \varphi}\delta_{jm}\!\int\!d^3\!zd^3\!y\,\Bigl[\mathcal D^{\varphi m}_{\mu}(A^{\ast}(\mathbf x))\mathscr A^{(\mu,y)}_{\;\;\;\;(\varepsilon,k\,z)}\Bigr]\\
&&\quad\quad\quad\quad\quad\quad\times\Bigl[\mathcal D^{\alpha i}_{\beta}(A^{\ast}(\mathbf x))d^{(\beta,x)(\epsilon',y)}\Bigr] \omega^{(\varepsilon,k,z)}\omega^{(\gamma,j,y)}
\end{eqnarray*}
$\mathbf{\Roman 1_{2)}}:$
\[
 c^{\varphi}_{\gamma \mu}\!\int\!d^3\!zd^3\!y\,\mathscr A^{(\mu,y)}_{\;\;\;\;(\varepsilon,k\,z)}\Bigl[\mathcal D^{\alpha i}_{\beta}(A^{\ast}(\mathbf x))\mathscr A^{(\beta,x)}_{\;\;\;\;(\varphi,j,\,y)}\Bigr]\omega^{(\varepsilon,k,z)}\omega^{(\gamma,j,y)}
\]
$\mathbf{\Roman 1_{3)}}:$
\[
-c^{\varphi}_{\gamma \mu} \delta_{j k}\!\int\!d^3\!z\,\Bigl[\mathcal D^{\alpha i}_{\beta}(A^{\ast}(\mathbf x))d^{(\beta,x)(\mu,z)}\Bigr]\omega^{(\varepsilon,k,z)}\omega^{(\gamma,j,z)}
\]
$\mathbf{\Roman 2}=-(\mathscr A^{\beta}_{M}K^A_{\beta,B}+\mathscr A^{\beta}_{B}K^A_{\beta,M})\,\omega^M\omega^B=-2\mathscr A^{\beta}_{M}K^A_{\beta,B},\omega^M\omega^B 
:$
\[
 -2\,c^{\alpha}_{\varepsilon \beta}\Bigl[\;\int\!d^3\!y\,\mathscr A^{(\beta,x)}_{\;\;\;\;(\gamma,j,\,y)}\,\omega^{(\gamma,j,y)}\Bigr]\omega^{(\varepsilon,i,x)}
\]
$\mathbf{\Roman 3}=\frac12(K^A_{\mu , D}K^D_{\sigma})(\mathscr A^{\mu}_{M}\mathscr A^{\sigma}_{B}+\mathscr A^{\sigma}_{M}\mathscr A^{\mu}_{B})\,\omega^M\omega^B=(K^A_{\mu , D}K^D_{\sigma})\mathscr A^{\mu}_{M}\mathscr A^{\sigma}_{B}\,\omega^M\omega^B$
\[
 c^{\alpha}_{\varphi \mu}\!\int\!d^3\!yd^3\!z\,\mathscr A^{(\mu,x)}_{\;\;\;\;(\gamma,j,\,y)}\Bigl[\mathcal D^{\varphi i}_{\nu}(A^{\ast}(\mathbf x))\mathscr A^{(\nu,x)}_{\;\;\;\;(\varepsilon,k,\,z)}\Bigr]\omega^{(\gamma,j,y)}\omega^{(\varepsilon,k,z)}
\]

${\Roman 3}$ can also be written in a following way:
\[
 c^{\alpha}_{\varphi \mu}\!\int\!d^3\!z
\Bigl[\mathcal D^{\varphi i}_{\nu}(A^{\ast}(\mathbf x))\mathscr A^{(\nu,x)}_{\;\;\;\;(\varepsilon,k,\,z)}\Bigr]\omega^{(\varepsilon,k,z)}\times\!\int\!d^3\!y\mathscr A^{(\mu,x)}_{\;\;\;\;(\gamma,j,\,y)}\omega^{(\gamma,j,y)}.
\]

${G^{AR}{ }^{\;\;\rm H}{\Gamma}_{BMR}}={\Roman 1_{1)}}+{\Roman 1_{2)}}+{\Roman 1_{3)}}+{\Roman 2}+{\Roman 3}.$

${}$

\begin{eqnarray*}
\mathbf{G^{AR}{ }^{\;\;\rm H}{\Gamma}_{BmR}}&=&-\frac12\bigl(\mathscr A^{\beta}_{B,m}K^A_{\beta}+\mathscr A^{\beta}_{m,B}K^A_{\beta}\bigr)
-\mathscr A^{\beta}_{m}K^A_{\beta,B}\\
&{}& +\frac12(K^A_{\mu, D}K^D_{\sigma})(\mathscr A^{\mu}_{m}\mathscr A^{\sigma}_{B}+\mathscr A^{\sigma}_{m}\mathscr A^{\mu}_{B})
\end{eqnarray*}

$G^{AR}{ }^{\;\;\rm H}{\Gamma}_{BmR}\,\omega^B\omega^m$, $A\to(\alpha,i,x)$, $B\to(\epsilon,k,z)$, $m\to(m,v)$\\

$G^{(\alpha,i,x)(...)}{ }^{\;\;\rm H}{\Gamma}_{(\epsilon,k,z)(m,v)(...)}\,\omega^{(\epsilon,k,z)}\omega^{(m,v)}$

\[
\mathscr A^{\beta}_{B,m}=-{d}^{\beta \varepsilon}\,K^r_{\varepsilon \,m}\,G_{rn}\,K^n_{\mu}{\mathcal A}^{\mu}_{B}-{\mathcal A}^{\beta}_{n}\,{\mathcal A}^{\mu}_{B}\,K^n_{\mu\, m}
\]
{}\\
$\mathbf{\Roman 1_{1)}}=\frac12({d}^{\beta \varepsilon}\,K^r_{\varepsilon \,m}\,G_{rn}\,K^n_{\mu}{\mathcal A}^{\mu}_{B})K^A_{\beta}\,\omega^B\omega^m:$

\[
 \frac12(\bar J_{\varepsilon'})^r_m G_{rn}(\bar J_{\mu})^n_{n'}\!\int\! d^3\!zd^3\!z'\Bigl[{\mathcal D}^{\alpha i}_{\beta}(A^{\ast}(\mathbf x))d^{(\beta,x)(\varepsilon',z')}\Bigr]\mathscr A^{(\mu,z')}_{\;\;\;\;(\varepsilon,k,z)}\tilde f^{n'}(\mathbf z')\omega^{(\varepsilon,k,z)}\omega^{(m,z')}.
\]\\
$\mathbf{\Roman 1_{2)}}=\frac12({\mathcal A}^{\beta}_{n}\,{\mathcal A}^{\mu}_{B}\,K^n_{\mu\, m})K^A_{\beta}\,\omega^B\omega^m:$

\[
 \frac12(\bar J_{\mu})^n_m \! \int\! d^3\!zd^3\!z'   \mathscr A^{(\mu,z')}_{\;\;\;\;(\varepsilon,k,z)}\Bigl[{\mathcal D}^{\alpha i}_{\beta}(A^{\ast}(\mathbf x))\mathscr A^{(\beta,x)}_{\;\;\;(n,z')}
\Bigr]\omega^{(\varepsilon,k,z)}\omega^{(m,z')}.
\]

\[
\mathscr A^{\beta}_{m,B}=
-{d}^{\beta \varepsilon}\,K^{A'}_{\varepsilon\, B}\,G_{A'B'}\,K^{B'}_{\mu}{\mathcal A}^{\mu}_{m}-{\mathcal A}^{\beta}_{D}\,{\mathcal A}^{\mu}_{m}\,K^D_{\mu\, B}.
\]
{}\\ 
$\mathbf{\Roman 1_{3)}}=\frac12({d}^{\beta \varepsilon}\,K^{A'}_{\varepsilon\, B}\,G_{A'B'}\,K^{B'}_{\mu}{\mathcal A}^{\mu}_{m})K^A_{\beta}\,\omega^B\omega^m:$
\[
 \frac12c^{\gamma}_{\varepsilon \varepsilon'}k_{\gamma \nu}\delta_{kk'}\!\int\! d^3\!zd^3\!z'\Bigl[{\mathcal D}^{\nu k'}_{\mu}(A^{\ast}(\mathbf z))\mathscr A^{(\mu,z)}_{\;\;\;(m,z')}\Bigr]\Bigl[{\mathcal D}^{\alpha i}_{\beta}(A^{\ast}(\mathbf x))d^{(\beta,x)(\varepsilon',z)}
\Bigr]
\omega^{(\varepsilon,k,z)}\omega^{(m,z')}.
\]
{}\\
$\mathbf{\Roman 1_{4)}}=\frac12\mathscr A^{\beta}_{m,B}K^A_{\beta}\,\omega^m \omega^B:$
\[
 \frac12c^{\gamma}_{\varepsilon \mu}\!\int\! d^3\!zd^3\!z'\mathscr A^{(\mu,z)}_{\;\;\;(m,z')}\Bigl[{\mathcal D}^{\alpha i}_{\beta}(A^{\ast}(\mathbf x))\mathscr A^{(\beta,x)}_{\;\;\;\;(\gamma,k,z)}\Bigr]\omega^{(\varepsilon,k,z)}\omega^{(m,z')}.
\]\\
The terms without dependence on $K^A_{\beta}$ are represented by\\ 
$(\mathbf{\Roman 2}+\mathbf{\Roman 3})\omega^B \omega^m :$
\begin{eqnarray*}
&&-\delta^i_k c^{\alpha}_{\epsilon \beta}\biggl(\int\! d^3v\,\mathscr A^{(\beta,x)}_{\;\;\;(m,v)}\omega^{(m,v)}\biggr)\omega^{(\epsilon,k,x)}\\
&&+\frac12c^{\alpha}_{\epsilon \beta}\int\!d^3v\,d^3z\biggl[\mathscr A^{(\beta,x)}_{\;\;\;\;(\epsilon,k,z)}\overleftarrow{\overrightarrow{\mathcal D}}{}^{\mu i}_{\nu}(A^{\ast}(\mathbf x))\mathscr A^{(\nu,x)}_{\;\;\;(m,v)}\omega^{(m,v)}
\omega^{(\epsilon,k,z)}\biggr]
\end{eqnarray*}

${G^{AR}{ }^{\;\;\rm H}{\Gamma}_{BmR}}\,\omega^B \omega^m =\Roman 1+{\Roman 2}+{\Roman 3}.$

${}$\\

$\mathbf{G^{AR}{ }^{\;\;\rm H}{\Gamma}_{pqR}}$

\begin{eqnarray*}
G^{AR}{ }^{\;\;\rm H}{\Gamma}_{pqR}&=&-\frac12\bigl(\mathscr A^{\beta}_{p,q}K^A_{\beta}+\mathscr A^{\beta}_{q,p}K^A_{\beta})\\
&{}& +\frac12(K^A_{\varepsilon, D}K^D_{\sigma})(\mathscr A^{\varepsilon}_{q}\mathscr A^{\sigma}_{p}+\mathscr A^{\sigma}_{q}\mathscr A^{\varepsilon}_{p})
\end{eqnarray*}
\[
\mathscr A^{\alpha}_{p,q}=-{d}^{\alpha \varepsilon}\,K^r_{\varepsilon\, q}\,G_{rn}\,K^n_{\mu}{\mathcal A}^{\mu}_{p}-{\mathcal A}^{\alpha}_{n}\,{\mathcal A}^{\mu}_{p}\,K^n_{\mu\, q}+{d}^{\alpha \mu}\,K^m_{\mu\, q}\,G_{mp}.
\]
$-\frac12(\mathscr A^{\beta}_{p,q}K^A_{\beta}+\mathscr A^{\beta}_{q,p}K^A_{\beta})\omega ^p\omega^q$ has three terms.\\
$\mathbf{\Roman 1_{1)}}:$
\begin{eqnarray*}
(\bar J_{\varepsilon})^{r'}_q G_{r'n'}(\bar J_{\mu})^{n'}_{m'}\!\int\! d^3\!yd^3\!z\tilde f^{m'}(\mathbf z)
\Bigl[{\mathcal D}^{\alpha i}_{\beta}(A^{\ast}(\mathbf x))d^{(\beta,x)(\varepsilon,z)}\Bigr]\mathscr A^{(\mu,z)}_{\;\;(p,y)}\omega^{(p,y)}\omega^{(q,z)}.
\end{eqnarray*}\\
$\mathbf{\Roman 1_{2)}}:$
\[
 (\bar J_{\mu})^{n}_{q}\!\int\! d^3\!yd^3\!z\mathscr A^{(\mu,z)}_{\;\;(p,y)}\Bigl[{\mathcal D}^{\alpha i}_{\beta}(A^{\ast}(\mathbf x))\mathscr A^{(\beta,x)}_{\;\;(n,z)}\Bigr]\omega^{(p,y)}\omega^{(q,z)}.
\]\\
$\mathbf{\Roman 1_{3)}}:$
\[
 -(\bar J_{\mu})^{n}_{q}G_{np}\!\int \!d^3\!y\Bigl[{\mathcal D}^{\alpha i}_{\beta}(A^{\ast}(\mathbf x))d^{(\beta,x)(\mu,y)}\Bigr]\omega^{(p,y)}\omega^{(q,y)}.
\]\\
$\mathbf{\Roman 2}=
+\frac12(K^A_{\beta, E}K^E_{\nu})(\mathscr A^{\beta}_{q}\mathscr A^{\nu}_{p}+\mathscr A^{\nu}_{q}\mathscr A^{\beta}_{p}):$
\[
 \frac12c^{\alpha}_{\mu \beta}\!\int\! d^3\!yd^3\!z\Bigl(A^{(\nu,x)}_{\;\;(p,y)}\overleftarrow{\overrightarrow{\mathcal D}}{}^{\,\mu i}_{\nu}(A^{\ast}(\mathbf x))\mathscr A^{(\beta,x)}_{\;\;(q,z)}\Bigr) \omega^{(p,y)}\omega^{(q,z)}.
\]
$G^{AR}{ }^{\;\;\rm H}{\Gamma}_{pqR}\,\omega^p\omega^q={\Roman 1_{1)}}+{\Roman 1_{2)}}+{\Roman 1_{3)}}+{\Roman 2}.$


${}$

$$\mathbf{G^{rm}{ }^{\;\;\rm H}{\Gamma}_{ABm}}=-\frac12\bigl(\mathscr A^{\beta}_{A,B}K^r_{\beta}+\mathscr A^{\beta}_{B,A}K^r_{\beta})
 +\frac12(K^r_{\mu, p}K^p_{\sigma})(\mathscr A^{\sigma}_{A}\mathscr A^{\mu}_{B}+\mathscr A^{\mu}_{A}\mathscr A^{\sigma}_{B})$$

$\mathscr A^{\beta}_{A,B}=-{d}^{\beta \mu}\,K^Q_{\mu, B}\,{\mathscr A}^{\epsilon}_{Q}\,K^D_{\epsilon}G_{DA}-{\mathscr A}^{\beta}_{Q}\,K^Q_{\nu , B}\,{\mathscr A}^{\nu}_{A}+{d}^{\beta \epsilon}\,K^D_{\epsilon,  B}\,G_{DA}$

$A\to(\alpha, n,z'),\;B\to(\gamma,k,z),\; \beta\to(\beta,u),\; \mu\to(\mu,v),\; Q\to(\nu,j,y),\;\\D\to(\sigma,m,y'),\;\epsilon\to(\epsilon,t),\;r\to(r,x).$ 

The terms of $\mathscr A^{\beta}_{A,B}$:\\
$\mathbf{\Roman 1_{1)}}:$
\[
 -\delta^j_k\, c^{\nu}_{\gamma \mu}\,d^{(\beta,u)(\mu,z)}\,k_{\sigma\alpha}\,\delta_{mn}\bigl[{\mathcal D}^{\sigma m}_{\epsilon}(A^{\ast}(\mathbf z'))\mathscr A^{(\epsilon,z')}_{\;\;\;\;(\nu,j,z)}\bigr]
\]\\
$\mathbf{\Roman 1_{2)}}:$
\[
 -c^{\nu}_{\gamma \mu}\mathscr A^{(\beta,u)}_{\;\;\;\;(\nu,k,z)}\mathscr A^{(\mu,z)}_{\;\;\;\;(\alpha,n,z')}
\]\\
$\mathbf{\Roman 1_{3)}}:$
\[
 c^{\sigma}_{\gamma \epsilon}\,k_{\sigma\alpha}\,\delta_{kn}\,d^{(\beta,u)(\epsilon, z)}\,\delta^3(\mathbf{z}-\mathbf{z'})
\]

$$\mathscr A^{\beta}_{B,A}=-{d}^{\beta \mu}\,K^Q_{\mu, A}\,{\mathscr A}^{\epsilon}_{Q}\,K^D_{\epsilon}G_{DB}-{\mathscr A}^{\beta}_{Q}\,K^Q_{\mu , A}\,{\mathscr A}^{\mu}_{B}+{d}^{\beta \epsilon}\,K^D_{\epsilon,  A}\,G_{DB}$$

The terms of $\mathscr A^{\beta}_{B,A}$:\\
$\mathbf{\Roman 1_{4)}}:$
\[
  -\delta^j_n\, c^{\nu}_{\alpha \mu}\,d^{(\beta,u)(\mu,z')}\,k_{\sigma\gamma}\,\delta_{mk}\bigl[{\mathcal D}^{\sigma m}_{\epsilon}(A^{\ast}(\mathbf z))\mathscr A^{(\epsilon,z)}_{\;\;\;\;(\nu,j,z')}\bigr]
\]\\
$\mathbf{\Roman 1_{5)}}:$
\[
 -c^{\nu}_{\alpha \mu}\mathscr A^{(\beta,u)}_{\;\;\;\;(\nu,n,z')}\mathscr A^{(\mu,z')}_{\;\;\;\;(\gamma,k,z)}
\]\\
$\mathbf{\Roman 1_{6)}}:$
\[
 c^{\sigma}_{\alpha \epsilon}\,k_{\sigma\gamma}\,\delta_{nk}\,d^{(\beta,u)(\epsilon, z')}\,\delta^3(\mathbf{z'}-\mathbf{z})
\]

(All terms of $\Roman 1$ must  be multiplied by  $K^r_{\beta}\,\omega^A\omega^B $.)\\
$\mathbf{\Roman 2}=\frac12(K^r_{\mu, p}K^p_{\sigma})(\mathscr A^{\sigma}_{A}\mathscr A^{\mu}_{B}+\mathscr A^{\mu}_{A}\mathscr A^{\sigma}_{B}):$
\[
 +\frac12(\bar J_{\mu})^r_{m'}(\bar J_{\epsilon})^{m'}_q \tilde f^q(\mathbf x)\Bigl(\mathscr A^{(\epsilon,x)}_{\;\;\;\;(\alpha,n,z')}
\mathscr A^{(\mu,x)}_{\;\;\;\;(\gamma,k,z)}+\mathscr A^{(\mu,x)}_{\;\;\;\;(\alpha,n,z')}
\mathscr A^{(\epsilon,x)}_{\;\;\;\;(\gamma,k,z)}\Bigr)
\]


${G^{rm}{ }^{\;\rm H}{\Gamma}_{ABm}}\,\omega^A\omega ^B={\Roman 1}+{\Roman 2}=$
\[
 + c^{\nu}_{\gamma \mu} (\bar J_{\beta})^r_{p}\tilde f^p(\mathbf x)\,k_{\sigma\alpha}\!\int\!d^3\!zd^3\!z'
d^{(\beta,x)(\mu,z)}\,\Bigl[{\mathcal D}^{\sigma }_{\epsilon n}(A^{\ast}(\mathbf z'))\mathscr A^{(\epsilon,z')}_{\;\;\;\;(\nu,k,z)}\Bigr]\omega^{(\alpha,n,z')}\omega^{(\gamma,k,z)}
\]
\[
 +c^{\nu}_{\gamma \mu}k_{\varphi \nu} (\bar J_{\beta})^r_{p}\tilde f^p(\mathbf x)\!\int\! d^3\!zd^3\!z'\mathscr A^{(\mu,z)}_{\;\;\;\;(\alpha,n,z')}\mathscr A^{(\beta,x)}_{\;\;\;\;(\nu,k,z)}\omega^{(\alpha,n,z')}\omega^{(\gamma,k,z)}
\]
 \[
  -c^{\sigma}_{\gamma \epsilon}\delta_{kn}k_{\sigma \alpha}(\bar J_{\beta})^r_{p}\,\tilde f^p(\mathbf x)\!\int \!d^3\!z\,d^{(\beta,x)(\epsilon, z)}\omega^{(\alpha,n,z)}\omega^{(\gamma,k,z)}
 \]
 \[
  +(\bar J_{\mu})^r_{m'}(\bar J_{\epsilon})^{m'}_q \tilde f^q(\mathbf x)\!\int\! d^3\!zd^3\!z'\mathscr A^{(\epsilon,x)}_{\;\;\;\;(\alpha,n,z')}
\mathscr A^{(\mu,x)}_{\;\;\;\;(\gamma,k,z)}\omega^{(\alpha,n,z')}\omega^{(\gamma,k,z)}
 \]
(Here the symmetry between $A$ and $B$ was taken into account.)

${}$

\begin{eqnarray*}
\mathbf{G^{rm}{ }^{\;\;\rm H}{\Gamma}_{pBm}}&=&-\frac12\bigl(\mathscr A^{\beta}_{B,p}+\mathscr A^{\beta}_{p,B})K^r_{\beta}-\mathscr A^{\beta}_{B}K^r_{\beta,p}\\
&{}& +\frac12(K^r_{\varepsilon, q}K^q_{\mu})(\mathscr A^{\mu}_{p}\mathscr A^{\varepsilon}_{B}+\mathscr A^{\mu}_{B}\mathscr A^{\varepsilon}_{p})
\end{eqnarray*}


$\mathscr A^{\beta}_{B,p}=-{d}^{\beta \varepsilon}\,K^{r'}_{\varepsilon \,p}\,G_{r'n}\,K^n_{\mu}{\mathscr A}^{\mu}_{B}-{\mathscr A}^{\beta}_{n}\,{\mathscr A}^{\mu}_{B}\,K^n_{\mu\, p}$

$B\to(\varepsilon,k,z)$, $p\to(p,z)$, $\beta\to(\beta,u)$\\

The terms of $-\frac12\mathbf{\mathscr A}{}^{\beta}_{B,p} K^r_{\beta}\,\omega^B\omega^p$:\\
$\mathbf{\Roman 1_{1)}}:$
\[
 \frac12(\bar J_{\varepsilon'})^{q}_{p}G_{q,n}(\bar J_{\mu})^{n}_{n'}\tilde f^{n'}\!(\mathbf x)(\bar J_{\beta})^{r}_{l}\tilde f^l(\mathbf x)\!\int\! d^3\!zd^3\!z'd^{(\beta,x)(\varepsilon',z')}\mathscr A^{(\mu,z')}_{\;\;\;(\varepsilon,k,z)}\omega^{(\varepsilon,k,z)}\omega^{(p,z')}
\]\\
$\mathbf{\Roman 1_{2)}}=-\frac12(-{\mathscr A}^{\beta}_{n}\,{\mathscr A}^{\mu}_{B}\,K^n_{\mu\,,p})K^r_{\beta}\,\omega^B\omega^p:$

\[
+\frac12(\bar J_{\mu})^{n}_{p} (\bar J_{\beta})^{r}_{q}\tilde f^q(\mathbf x)\!\int\! d^3\!zd^3\!z'  \mathscr A^{(\beta,x)}_{\;\;(n,z')}            \mathscr A^{(\mu,z')}_{\;\;\;(\varepsilon,k,z)}\omega^{(\varepsilon,k,z)}\omega^{(p,z')}
\]


$\mathscr A^{\beta}_{p,B}=-{d}^{\beta \varepsilon}\,K^{A'}_{\varepsilon, \,B}\,G_{A'B'}\,K^{B'}_{\mu}{\mathscr A}^{\mu}_{p}-{\mathscr A}^{\beta}_{D}\,{\mathscr A}^{\mu}_{p}\,K^D_{\mu\,, B}$

The terms of $-\frac12\mathbf{\mathscr A}{}^{\beta}_{p,B}K^r_{\beta}\,\omega^B\omega^p$:\\
$\mathbf{\Roman 1_{3)}}:$

\[
 +\frac12c^{\gamma}_{\varepsilon\varepsilon'}k_{\gamma \nu}(\bar J_{\beta})^{r}_{q}\tilde f^q(\mathbf x)\delta_{kk'}\!\int\! d^3\!zd^3\!z'd^{(\beta,x)(\varepsilon',z)}\Bigl[\mathcal D^{\nu k'}_{\mu}(A^{\ast}(z))\mathscr A^{(\mu,z)}_{\;\;(p,z')}\Bigr]\omega^{(\varepsilon,k,z)}\omega^{(p,z')}
\]
$\mathbf{\Roman 1_{4)}}=-\frac12(-{\mathscr A}^{\beta}_{D}\,{\mathscr A}^{\mu}_{p}\,K^D_{\mu\,, B})K^r_{\beta}\,\omega^B\omega^p:$
\[
 +\frac12c^{\gamma}_{\varepsilon \mu}(\bar J_{\beta})^{r}_{q}\tilde f^q(\mathbf x)\!\int\! d^3\!zd^3\!z'\mathscr A^{(\beta,x)}_{\;\;\;(\gamma,k,z)}\mathscr A^{(\mu,z)}_{\;\;\;(p,z')}\,\omega^{(\varepsilon,k,z)}\omega^{(p,z')}\\
\]
$\mathbf{\Roman 2}=-\mathscr A^{\beta}_{B}K^r_{\beta,p}\omega^B\omega^p$:
\[
 -(\bar J_{\beta})^{r}_{p}\Bigl[\int \!d^3\!z \,\mathscr A^{(\beta,x)}_{\,\;\;\;(\varepsilon,k,z)}\,\omega^{(\varepsilon,k,z)}\Bigr]  \omega^{(p,x)}           
\]\\
$\mathbf{\Roman 3}= \frac12(K^r_{\varepsilon, q}K^q_{\mu})(\mathscr A^{\mu}_{p}\mathscr A^{\varepsilon}_{B}+\mathscr A^{\mu}_{B}\mathscr A^{\varepsilon}_{p})\omega^B\omega^p:$
\[
\frac12(\bar J_{\gamma})^{r}_{k}(\bar J_{\mu})^{k}_{n}\tilde f^n(\mathbf x)\!\int\! d^3\!zd^3\!z'\Bigl( \mathscr A^{(\mu,x)}_{\;\;\;(p,z')}\mathscr A^{(\gamma,x)}_{\,\;\;\;(\varepsilon,k,z)}+\mathscr A^{(\gamma,x)}_{\;\;\;(p,z')}\mathscr A^{(\mu,x)}_{\,\;\;\;(\varepsilon,k,z)}\Bigr)\omega^{(\varepsilon,k,z)}\omega^{(p,z')}
\]
${G^{rm}{ }^{\;\;\rm H}{\Gamma}_{pBm}}\,\omega^B\omega^p={\Roman 1_{1)}}+{\Roman 1_{2)}}+{\Roman 1_{3)}}+{\Roman 1_{4)}}+{\Roman 2}+{\Roman 3}$\\

${}$

\begin{eqnarray*}
\mathbf{G^{rm}{ }^{\;\;\rm H}{\Gamma}_{pqm}}&=&-\frac12\bigl(\mathscr A^{\beta}_{p,q}K^r_{\beta}+\mathscr A^{\beta}_{q,p}K^r_{\beta})
-(\mathscr A^{\beta}_{p}K^r_{\beta,q}+\mathscr A^{\beta}_{q}K^r_{\beta,p})\\
&{}& +\frac12(K^r_{\mu, n}K^n_{\nu})(\mathscr A^{\mu}_{q}\mathscr A^{\nu}_{p}+\mathscr A^{\nu}_{q}\mathscr A^{\mu}_{p})
\end{eqnarray*}

$$\mathscr A^{\beta}_{p,q}=-{d}^{\beta \varepsilon}\,K^r_{\varepsilon\, q}\,G_{rn}\,K^n_{\mu}{\mathscr A}^{\mu}_{p}-{\mathscr A}^{\beta}_{n}\,{\mathscr A}^{\mu}_{p}\,K^n_{\mu\, q}+{d}^{\beta \mu}\,K^m_{\mu\, q}\,G_{mp}$$

$\mathbf{\Roman 1_{1)}}=-\frac12\,2\,\mathscr A^{\beta}_{p,q}K^r_{\beta}\,\omega^{p}\omega^{q}:$

\[
 (\bar J_{\beta})^{r}_{n}\tilde f^{n}(\mathbf x)(\bar J_{\varepsilon})^{r'}_{q}G_{r'n'}(\bar J_{\mu})^{n'}_{m'}\!\int\! d^3\!zd^3\!y\,d^{(\beta,x)(\varepsilon,z)}\tilde f^{m'}(\mathbf z)\,\mathscr A^{(\mu,z)}_{\;\;\;(p,y)}\,\omega^{(p,y)}\omega^{(q,z)}
\]

$\mathbf{\Roman 1_{2)}}=-\frac12\,2\,(-{\mathscr A}^{\beta}_{n}\,{\mathscr A}^{\mu}_{p}\,K^n_{\mu\, q}K^r_{\beta})\omega^{p}\omega^{q}:$
\[
 (\bar J_{\mu})^{n}_{q}(\bar J_{\beta})^{r}_{n'}\tilde f^{n'}(\mathbf x)\!\int\!d^3\!zd^3\!y\,\mathscr A^{(\beta,x)}_{\;\;\;(n,z)}\mathscr A^{(\mu,z)}_{\;\;\;(p,y)}\,\omega^{(p,y)}\omega^{(q,z)}
\]

$\mathbf{\Roman 1_{3)}}=-\frac12\,2\,(+{d}^{\beta \mu}\,K^m_{\mu\, q}\,G_{mp}K^r_{\beta})\omega^{p}\omega^{q}:$
\[
 -(\bar J_{\mu})^{n}_{q}G_{np}(\bar J_{\beta})^{r}_{n'}\tilde f^{n'}(\mathbf x)\!\int\!d^3\!y\, d^{(\beta,x)(\mu,y)}\,\omega^{(p,y)}\omega^{(q,y)}
\]

$\mathbf{\Roman 2}=2\,(-\mathscr A^{\beta}_{p}K^r_{\beta,q})\omega^{p}\omega^{q}:$
\[
 -2(\bar J_{\beta})^{r}_{q}\biggl[\;\int\!d^3\!y\,\mathscr A^{(\beta,z)}_{\;\;\;(p,y)}\,\omega^{(p,y)}\biggr]\omega^{(q,x)}
\]

$\mathbf{\Roman 3}=2\frac12(K^r_{\mu, n}K^n_{\nu})\,\mathscr A^{\mu}_{q}\mathscr A^{\nu}_{p}\,\omega^{q}\omega^{p}:$
\[
 (\bar J_{\mu})^{r}_{m}(\bar J_{\nu})^{m}_{k}\tilde f^{k}(\mathbf x)\!\int\!d^3\!z\,d^3\!y\,\mathscr A^{(\mu,x)}_{\;\;\;(q,z)}\mathscr A^{(\nu,x)}_{\;\;\;(p,y)}\,\omega^{(p,y)}\omega^{(q,z)}
\]

${\Roman 3}$ can also be written in a following way:
\[
\Bigl((\bar J_{\nu})^{m}_{k}\tilde f^{k}(\mathbf x)\!\int\!d^3\!y\,A^{(\nu,x)}_{\;\;\;(p,y)}\,\omega^{(p,y)}\Bigr) \Bigl((\bar J_{\mu})^{r}_{m}\!\int\!d^3\!z\,\mathscr A^{(\mu,x)}_{\;\;\;(q,z)}\omega^{(q,z)}\Bigr)
\]

${G^{rm}{ }^{\;\;\rm H}{\Gamma}_{pqm}}={\Roman 1_{1)}}+{\Roman 1_{2)}}+{\Roman 1_{3)}}+{\Roman 2}+{\Roman 3}$

${}$

$$\mathbf{G^{AR}{\mathscr F}^{\alpha'}_{QR}\,\omega^Q\,p_{\alpha'}}$$

\begin{eqnarray*}
&&G^{AR}{\mathscr F}^{\alpha'}_{QR}=-(K^S_{\varphi,Q})(d^{\varphi \alpha'}\mathscr A^{\mu}_S+d^{\varphi \mu}\mathscr A^{\alpha'}_S)K^A_{\mu}-(K^A_{\epsilon ,B }K^B_{\nu})({d}^{\alpha' \epsilon}{\mathscr A}^{\nu}_{Q}+{d}^{\alpha' \nu }{\mathscr A}^{\epsilon}_{Q})
\nonumber\\
&&+2{d}^{\,\alpha' \mu}K^A_{\mu, Q}+c^{\alpha'}_{\nu\mu}{d}^{\mu\varphi}{\mathscr A}^{\nu}_{Q}K^A_{\varphi}
\label{7}
\end{eqnarray*}

$A=(\alpha,i,x),\quad Q=(\varepsilon,k,z)$

$\mathbf{\Roman 1_{1)}}=-(K^S_{\varphi,Q})d^{\varphi \alpha'}\mathscr A^{\mu}_SK^A_{\mu}\,\omega^Q\,p_{\alpha'}:$
\[
-c^{\gamma}_{\varepsilon\varphi} \!\int\!d^3\!u\,d^3\!z\,d^{(\varphi ,z)(\alpha',u)}\Bigl[\mathcal D^{\alpha i}_{\mu}(A^{\ast}(x))\mathscr A^{(\mu,x)}_{\;\;\;\;(\gamma,k,z)}\Bigr]\omega^{(\varepsilon,k,z)}\,p_{(\alpha',u)}
\]

$\mathbf{\Roman 1_{2)}}=-(K^S_{\varphi,Q})d^{\varphi \mu}\mathscr A^{\alpha'}_SK^A_{\mu}\,\omega^Q\,p_{\alpha'}:$ 

\[
-c^{\gamma}_{\varepsilon\varphi} \!\int\!d^3\!u\,d^3\!z\,\mathscr A^{(\alpha',u)}_{\;\;\;\;(\gamma,k,z)}     \Bigl[\mathcal D^{\alpha i}_{\mu}(A^{\ast}(x))d^{(\varphi ,z)(\mu,x)}\Bigr]\omega^{(\varepsilon,k,z)}\,p_{(\alpha',u)}
\]

$\mathbf{\Roman 2_{1)}}=-(K^A_{\epsilon ,B }K^B_{\nu}){d}^{\alpha' \epsilon}{\mathscr A}^{\nu}_{Q}\,\omega^Q\,p_{\alpha'}:$

\[
 -c^{\alpha}_{\mu'\varphi}\!\int\!d^3\!u\,d^3\!z\,d^{(\alpha' ,u)(\varphi,x)}\Bigl[\mathcal D^{\mu' i}_{\nu}(A^{\ast}(x))
\mathscr A^{(\nu,x)}_{\;\;\;\;(\varepsilon,k,z)}\Bigr]\omega^{(\varepsilon,k,z)}\,p_{(\alpha',u)}
\]

$\mathbf{\Roman 2_{2)}}=-(K^A_{\epsilon ,B }K^B_{\nu}){d}^{\alpha' \nu }{\mathscr A}^{\epsilon}_{Q}\,\omega^Q\,p_{\alpha'}:$

\[
 -c^{\alpha}_{\mu'\varphi}\!\int\!d^3\!u\,d^3\!z\,\mathscr A^{(\varphi,x)}_{\;\;\;\;(\varepsilon,k,z)}
\Bigl[\mathcal D^{\mu' i}_{\nu}(A^{\ast}(x))d^{(\alpha' ,u)(\nu,x)}\Bigr]\omega^{(\varepsilon,k,z)}\,p_{(\alpha',u)}
\]

$\mathbf{\Roman 3}=$ $2{d}^{\,\alpha' \mu}K^A_{\mu, Q}:$

\[
 2c^{\alpha}_{\varepsilon \mu}\Bigl[\;\;\int\!d^3\!u\,d^{(\beta ,u)(\mu,x)}p_{(\beta,u)}\Bigr]\omega^{(\varepsilon,i,x)}
\]

$\mathbf{\Roman 4}=$ $c^{\alpha'}_{\nu\mu}{d}^{\mu\varphi}{\mathscr A}^{\nu}_{Q}K^A_{\varphi},$

$$\rm{where}\quad c^{(\alpha',u)}_{(\nu,y)(\mu,v)}=c^{\alpha'}_{\nu\mu}\delta^3(\mathbf u-\mathbf y)\delta^3(\mathbf u-\mathbf v)\,:$$

\[
 c^{\alpha'}_{\nu\mu}\!\int\!d^3\!u\,d^3\!z\,\mathscr A^{(\nu,u)}_{\;\;\;\;(\varepsilon,k,z)}\Bigl[\mathcal D^{\alpha i}_{\varphi}(A^{\ast}(x))d^{(\mu ,u)(\varphi,x)}\Bigr]\omega^{(\varepsilon,k,z)}\,p_{(\alpha',u)}
\]\\
($K^A_{\mu}$-terms:  first  ($\Roman 1$) and  fourth  ($\Roman 4$)).\\

$G^{AR}{\mathscr F}^{\alpha'}_{QR}\,\omega^Q\,p_{\alpha'}=\Roman 1+\Roman 2+\Roman 3+\Roman 4.$

${}$

$$\mathbf{G^{AR}{\mathscr F}^{\alpha'}_{q\,R}\,\omega^q\,p_{\alpha'}}$$

\begin{eqnarray*}
G^{AR}{\mathcal F}^{\alpha'}_{qR}&=&-(K^r_{\mu,q})(d^{\mu \alpha'}\mathscr A^{\varphi}_r+d^{\mu\varphi}\mathscr A^{\alpha'}_r)K^A_{\varphi}-(K^A_{\nu ,B }K^B_{\varphi})({d}^{\alpha' \nu}{\mathscr A}^{\varphi}_{q}+{d}^{\alpha' \varphi }{\mathscr A}^{\nu}_{q})
\nonumber\\
&{ }&+c^{\alpha'}_{\nu\mu}{d}^{\mu\varphi}{\mathscr A}^{\nu}_{q}K^A_{\varphi}
\end{eqnarray*}

$A=(\alpha,i,x),\quad q=(q,y),\quad \alpha'=(\alpha',u)$

$\mathbf{\Roman 1_{1)}}=-(K^r_{\mu,q})d^{\mu \alpha'}\mathscr A^{\varphi}_r K^A_{\varphi}\omega^q\,p_{\alpha'}:$

\[
 -(\bar J_{\mu})^{k}_{q}\!\int\!d^3\!u\,d^3\!y\,d^{(\mu,y)(\alpha',u)}\Bigl[\mathcal D^{\alpha i}_{\varphi}(A^{\ast}(x))
\mathscr A^{(\varphi,x)}_{\;\;\;\;(k,y)}\Bigr]\,\omega^{(q,y)}p_{(\alpha',u)}
\]

$\mathbf{\Roman 1_{2)}}=-(K^r_{\mu,q})d^{\mu\varphi}\mathscr A^{\alpha'}_rK^A_{\varphi}\,\omega^q\,p_{\alpha'}:$

\[
-(\bar J_{\mu})^{k}_{q}\!\int\!d^3\!u\,d^3\!y\,\mathscr A^{(\alpha',u)}_{\;\;\;\;(k,y)}\Bigl[\mathcal D^{\alpha i}_{\varphi}(A^{\ast}(x))d^{(\mu,y)(\varphi,x)}\Bigr]\,\omega^{(q,y)}p_{(\alpha',u)}
\]

$\mathbf{\Roman 2_{1)}}=-(K^A_{\nu ,B }K^B_{\varphi}){d}^{\alpha' \nu}{\mathscr A}^{\varphi}_{q}\,\omega^{(q,y)}p_{(\alpha',u)}:$

\[
-c^{\alpha}_{\mu' \nu}\!\int\!d^3\!u\,d^3\!y\,d^{(\alpha',u)(\nu,x)}\Bigl[\mathcal D^{\mu' i}_{\varphi}(A^{\ast}(x))\mathscr A^{(\varphi,x)}_{\;\;\;\;(q,y)}\Bigr]\,\omega^{(q,y)}p_{(\alpha',u)}
\]

$\mathbf{\Roman 2_{2)}}= -(K^A_{\nu ,B }K^B_{\varphi}) d^{\alpha' \varphi }{\mathscr A}^{\nu}_{q})\,\omega^{(q,y)}p_{(\alpha',u)}:$

\[
-c^{\alpha}_{\mu' \nu}\!\int\!d^3\!u\,d^3\!y\,\mathscr A^{(\nu,x)}_{\;\;\;\;(q,y)}\Bigl[\mathcal D^{\mu' i}_{\varphi}(A^{\ast}(x))d^{(\alpha',u)(\varphi,x)}\Bigr]\,\omega^{(q,y)}p_{(\alpha',u)}
\]

$\mathbf{\Roman 3}=c^{\alpha'}_{\nu\mu}{d}^{\mu\varphi}{\mathscr A}^{\nu}_{q}K^A_{\varphi}\,\omega^{(q,y)}p_{(\alpha',u)}:$
\[
c^{\alpha'}_{\nu \mu}\!\int\!d^3\!u\,d^3\!y\,\mathscr A^{(\nu,u)}_{\;\;\;\;(q,y)}\Bigl[\mathcal D^{\alpha i}_{\varphi}(A^{\ast}(x))d^{(\mu,u)(\varphi,x)}\Bigr]\,\omega^{(q,y)}p_{(\alpha',u)}
\]

($K^A_{\mu}$-terms:  first  ($\Roman 1$) and  third  ($\Roman 3$)).\\

$G^{AR}{\mathscr F}^{\alpha'}_{q\,R}\,\omega^Q\,p_{\alpha'}={\Roman 1}+{\Roman 2}+{\Roman 3}$

${}$

$$\mathbf{G^{rm}{\mathscr F}^{\alpha'}_{Q\,m}\,\omega^Q\,p_{\alpha'}}$$

\begin{eqnarray*}
 G^{rm}{\mathscr F}^{\alpha'}_{Q\,m}&=&-K^T_{\mu, Q}({d}^{\alpha' \mu}{\mathscr A}^{\beta}_{T}+{d}^{\beta \mu}{\mathscr A}^{\alpha'}_{T})K^r_{\beta}
-(K^n_{\nu }K^r_{\mu,n})\,({d}^{\alpha' \mu}{\mathscr A}^{\nu}_{Q}+{d}^{\alpha' \nu}{\mathscr A}^{\mu}_{Q})
\nonumber\\
&{}&+c^{\alpha'}_{\nu \mu}\,{d}^{\mu \beta}{\mathscr A}^{\nu}_{Q}\,K^r_{\beta}.
\end{eqnarray*}

$r=(r,x),\quad Q=(\varepsilon,k,z),\quad \alpha'=(\alpha',u)$\\
$\mathbf{\Roman 1_{1)}}=-K^T_{\mu, Q}{d}^{\alpha' \mu}{\mathscr A}^{\beta}_{T}\,\omega^Q p_{\alpha'}:$\\
\[
-c^{\gamma}_{\varepsilon \mu}(\bar J_{\beta})^{r}_{m}\tilde f^m(\mathbf x)\!\int\!d^3\!u\,d^3\!z\,d^{(\alpha',u)(\mu,z)}\,\mathscr A^{(\beta,x)}_{\;\;\;\;(\gamma,k,z)}\,\omega^{(\varepsilon,k,z)}p_{(\alpha',u)}
\]\\
$\mathbf{\Roman 1_{2)}}=-K^T_{\mu, Q}{d}^{\beta \mu}{\mathscr A}^{\alpha'}_{T}K^r_{\beta}\,\omega^Q p_{\alpha'}:$\\
\[
-c^{\gamma}_{\varepsilon \mu}(\bar J_{\beta})^{r}_{m}\tilde f^m(\mathbf x)\!\int\!d^3\!u\,d^3\!z\,d^{(\beta,x)(\mu,z)}\,\mathscr A^{(\alpha',u)}_{\;\;\;\;(\gamma,k,z)}\,\omega^{(\varepsilon,k,z)}p_{(\alpha',u)}
\]\\
$\mathbf{\Roman 2_{1)}}=-(K^n_{\nu }K^r_{\mu,n})\,{d}^{\alpha' \mu}{\mathscr A}^{\nu}_{Q}\,\omega^Qp^{\alpha'}:$
\[
-(\bar J_{\mu})^{r}_{m'}(\bar J_{\nu})^{m'}_{q}\tilde f^q(\mathbf x)\!\int\!d^3\!u\,d^3\!z\,d^{(\alpha',u)(\mu,x)}\,\mathscr A^{(\nu,x)}_{\;\;\;\;(\varepsilon,k,z)}\,\omega^{(\varepsilon,k,z)}p_{(\alpha',u)}
\]\\
$\mathbf{\Roman 2_{2)}}=-(K^n_{\nu }K^r_{\mu,n})\,{d}^{\alpha' \nu}{\mathscr A}^{\mu}_{Q}\,\omega^Qp^{\alpha'}:$
\[
-(\bar J_{\mu})^{r}_{m'}(\bar J_{\nu})^{m'}_{q}\tilde f^q(\mathbf x)\!\int\!d^3\!u\,d^3\!z\,d^{(\alpha',u)(\nu,x)}\,\mathscr A^{(\mu,x)}_{\;\;\;\;(\varepsilon,k,z)}\,\omega^{(\varepsilon,k,z)}p_{(\alpha',u)}
\]\\
$\mathbf{\Roman 3}=c^{\alpha'}_{\nu \mu}\,{d}^{\mu \beta}{\mathscr A}^{\nu}_{Q}\,K^r_{\beta}\,\omega^Qp^{\alpha'}:$

\[
c^{\alpha'}_{\nu \mu}(\bar J_{\beta})^{r}_{n}\tilde f^n(\mathbf x)\!\int\!d^3\!u\,d^3\!z\,d^{(\mu,u)(\beta,x)}\,\mathscr A^{(\nu,u)}_{\;\;\;\;(\varepsilon,k,z)}\,\omega^{(\varepsilon,k,z)}p_{(\alpha',u)}
\]

$G^{rm}{\mathscr F}^{\alpha'}_{Q\,m}\omega^Q p_{\alpha'}={\Roman 1}+{\Roman 2}+{\Roman 3}$.

${}$

$$\mathbf{G^{rm}{\mathscr F}^{\alpha'}_{q\,m}\,\omega^q\,p_{\alpha'}}$$

\begin{eqnarray*}
 G^{rm}{\mathcal F}^{\alpha'}_{q\,m}&=&-K^n_{\mu, q}({d}^{\alpha' \mu}{\mathscr A}^{\beta}_{n}+{d}^{\beta \mu}{\mathscr A}^{\alpha'}_{n})K^r_{\beta}
-(K^p_{\nu }K^r_{\mu,p})\,({d}^{\,\alpha' \mu}{\mathscr A}^{\nu}_{q}+{d}^{\,\alpha' \nu}{\mathscr A}^{\mu}_{q})
\nonumber\\
&{}&+2{d}^{\,\alpha' \beta} K^r_{\beta, q}   +c^{\alpha'}_{\nu \mu}\,{d}^{\mu \beta}{\mathscr A}^{\nu}_{q}\,K^r_{\beta}.
\end{eqnarray*}

$r=(r,x),\quad q=(q,z),\quad \alpha'=(\alpha',u)$\\

$\mathbf{\Roman 1_{1)}}=-K^n_{\mu, q}{d}^{\alpha' \mu}{\mathscr A}^{\beta}_{n}K^r_{\beta}\,\omega^q\,p_{\alpha'}:$

\[
-(\bar J_{\mu})^{n}_{q}(\bar J_{\beta})^{r}_{m}\tilde f^m(\mathbf x)\!\int\!d^3\!u\,d^3\!z\,\mathscr A^{(\beta,x)}_{\;\;\;(n,z)}d^{(\alpha',u)(\mu,z)}\,\,\omega^{(q,z)}p_{(\alpha',u)}
\]

$\mathbf{\Roman 1_{2)}}=-K^n_{\mu, q}{d}^{\beta \mu}{\mathscr A}^{\alpha'}_{n}K^r_{\beta}\,\omega^q\,p_{\alpha'}:$

\[
-(\bar J_{\mu})^{n}_{q}(\bar J_{\beta})^{r}_{m}\tilde f^m(\mathbf x)\!\int\!d^3\!u\,d^3\!z\,d^{(\beta,x)(\mu,z)}\mathscr A^{(\alpha',u)}_{\;\;\;\;(n,z)}\,\,\omega^{(q,z)}p_{(\alpha',u)}
\]

$\mathbf{\Roman 2_{1)}}=-(K^r_{\mu,p }K^p_{\nu})\,{d}^{\alpha' \mu}{\mathscr A}^{\nu}_{q}\,\omega^q\,p_{\alpha'}:$

\[
-(\bar J_{\mu})^{r}_{m'}(\bar J_{\nu})^{m'}_{n'}\tilde f^{n'}(\mathbf x)\!\int\!d^3\!u\,d^3\!z\,d^{(\alpha',u)(\mu,x)}\mathscr A^{(\nu,x)}_{\;\;\;\;(q,z)}\,\,\omega^{(q,z)}p_{(\alpha',u)}
\]

$\mathbf{\Roman 2_{2)}}=-(K^r_{\mu,p }K^p_{\nu})\,{d}^{\alpha' \nu}{\mathscr A}^{\mu}_{q}\,\omega^q\,p_{\alpha'}:$

\[
-(\bar J_{\mu})^{r}_{m'}(\bar J_{\nu})^{m'}_{n'}\tilde f^{n'}(\mathbf x)\!\int\!d^3\!u\,d^3\!z\,d^{(\alpha',u)(\nu,x)}\mathscr A^{(\mu,x)}_{\;\;\;\;(q,z)}\,\,\omega^{(q,z)}p_{(\alpha',u)}
\]

$\mathbf{\Roman 3}=2\,{d}^{\alpha' \beta}\,K^r_{\beta,q }\,\omega^q\,p_{\alpha'}:$

\[
2\,(\bar J_{\beta})^{r}_{q}\,\omega^{(q,x)}\!\int\!d^3\!u\,\,d^{(\alpha',u)(\beta,x)}\,p_{(\alpha',u)}
\]

$\mathbf{\Roman 4}=c^{\alpha'}_{\nu \mu}\,{d}^{\mu \beta}{\mathscr A}^{\nu}_{Q}\,K^r_{\beta}\,\omega^q\,p_{\alpha'}:$

\[
c^{\alpha'}_{\nu \mu}\,(\bar J_{\beta})^{r}_{m}\tilde f^{m}(\mathbf x)\!\int\!d^3\!u\,d^3\!z\,d^{(\mu,u)(\beta,x)}\mathscr A^{(\nu,x)}_{\;\;\;\;(q,z)}\,\,\omega^{(q,z)}p_{(\alpha',u)}
\]

$G^{rm}{\mathcal F}^{\alpha'}_{q\,m}\,\omega^q\,p_{\alpha'}={\Roman 1}+{\Roman 2}+{\Roman 3}+{\Roman 4}.$

${}$

$$\mathbf{G^{AR}({\mathscr D}_R\,d^{\kappa\sigma})p_{\kappa}p_{\sigma}}$$

\[
 G^{AR}({\mathscr D}_Rd^{\kappa\sigma})p_{\kappa}p_{\sigma}=2\,\Bigl[(K^D_{\beta}K^A_{\mu,D})\,d^{\beta\kappa}d^{\mu\sigma}+c^{\kappa}_{\beta \mu}d^{\beta\epsilon}d^{\mu\sigma}K^A_{\epsilon}\Bigr]p_{\kappa}p_{\sigma}
\]

$A=(\alpha,i,x),\quad \beta=(\beta,v),\quad \mu=(\mu,u),\quad \kappa=(\kappa,z),\quad \sigma=(\sigma,z')$

\[
\mathbf{\Roman 1}= 2\,c^{\alpha}_{\mu' \mu}\!\int\!d^3\!z\,d^3\!z'\,d^{(\mu,x)(\sigma,z')}\Bigl[\mathcal D^{\mu' i}_{\beta}(A^{\ast}(x))d^{(\beta,x)(\kappa,z)}\Bigr]p_{(\kappa,z)}p_{(\sigma,z')}
\]

\[
\mathbf{\Roman 2}= 2\, c^{\kappa}_{\beta\mu}\!\int\!d^3\!z\,d^3\!z'\,d^{(\mu,z)(\sigma,z')}\Bigl[\mathcal D^{\alpha i}_{\varepsilon}(A^{\ast}(x))d^{(\beta,z)(\varepsilon,x)}\Bigr]p_{(\kappa,z)}p_{(\sigma,z')}
\]

${}$

$$\mathbf{G^{rm}({\mathscr D}_md^{\kappa\sigma})p_{\kappa}p_{\sigma}}$$

\[
 G^{rm}({\mathscr D}_md^{\kappa\sigma})p_{\kappa}p_{\sigma}=2\,\Bigl[(K^n_{\beta}K^r_{\mu,n})\,d^{\beta\kappa}d^{\mu\sigma}+c^{\kappa}_{\beta \mu}d^{\beta\epsilon}d^{\mu\sigma}K^r_{\epsilon}\Bigr]p_{\kappa}p_{\sigma}
\]

\[
 \mathbf{\Roman 1}=2\,(\bar J_{\mu})^{r}_{n}(\bar J_{\beta})^{n}_{q}\tilde f^{q}(\mathbf x)\!\int\!d^3\!z\,d^3\!z'\,d^{(\beta,x)(\kappa,z)}d^{(\mu,x)(\sigma,z')}p_{(\kappa,z)}p_{(\sigma,z')}
\]

\[
 \mathbf{\Roman 2}=2\,c^{\kappa}_{\beta \mu}\,(\bar J_{\varepsilon})^{r}_{n}\tilde f^{n}(\mathbf x)\!\int\!d^3\!z\,d^3\!z'\,d^{(\beta,z)(\varepsilon,x)}d^{(\mu,z)(\sigma,z')}p_{(\kappa,z)}p_{(\sigma,z')}
\]


\begin{thebibliography}{**}
\bibitem{AbrMarsd}
R. Abraham and J. E. Marsden, {\it Foundation of Mechanics, 2nd Ed.}
(Addison-Wesley Redwood City, 1985).


\bibitem{Marsden}
J. E. Marsden, {\it Lecture on Mechanics}, London Math. Soc. Lect. Notes Series  174, (Cambridge University press, Cambridge, 1992).

\bibitem{Narasimhan}
M. S. Narasimhan, T. R. Ramadas, Geometry of $SU(2)$ Gauge Fields, {\it Commun. Math. Phys.} {\bf 67} (1979), 121-136. 


\bibitem{Daniel-Viallet}
M. Daniel and C. M. Viallet, Geometrical setting of Yang--Mills gauge theories, {\it Rev. Mod. Phys} {\bf 52} (1980), 175-197.

\bibitem{Mitter-Viallet}
P. K. Mitter and C. M. Vialette, On the bundle of connections and the gauge orbit manifold in Yang-Mills theory, {\it Comm. Math. Phys.} {\bf 79}  (1981), 457-472.

\bibitem{Babelon-Viallet}
O. Babelon and C. M. Viallet, The riemannian geometry of the configuration space of gauge thories
 {\it Comm. Math. Phys.} {\bf 81}  (1981), 515-525.


\bibitem{Creutz}
M. Creutz, I. J. Muzinich, and T. N. Tudron, Gauge fixing and canonical quantization, 
{\it Phys. Rev.} {\bf D19} no. 2,  (1979), 531-539.

\bibitem{Gawedzki}
K. Gawedzki, Yang-Mills theory as Schr\"{o}dinger quantum mechanics on the space of gauge-group orbit, {\it Phys. Rev.} {\bf D26}  (1982), 3593-3609.

\bibitem{Huffel-Kelnhofer}
H. H\"{u}ffel and G. Kelnhofer, QED revisited: proving equivalence between path integral and stochastic quantization, {\it Phys. Lett.} {\bf B588}  (2004), 145-150.

\bibitem{Stor_lagr_poinc_1}
S.N.Storchak, The Lagrange-Poincaré equations for a mechanical system with symmetry  on the principal fiber bundle 
over the base represented by the bundle space of the  associated bundle, arXiv: 1612.08897.

\bibitem{Stor_lagr_poinc_2}
S.N.Storchak, Coordinate representation of the Lagrange-Poincar\'{e} equations for a mechanical system with symmetry on the total space of a principal fiber bundle whose base is the bundle space of the associated bundle, arXiv: 1709.09030.

\bibitem{Razumov}
O. A. Khrustalev,A. V. Razumov, A. Yu. Taranov, {\it Nucl. Phys.}
{\bf B172} (1980) 44;\\
A. V. Razumov, A. Yu. Taranov,
{\it Teor. i Mat. Fyz.} {\bf 52} (1982) 34 (in Russian);\\
IHEP Preprint 82--41, Serpukhov, 1982;\\
A. V. Razumov, Bogolubov Transformation and Quantum Theory of
Constrained Systems, {\it Dissertation} (Protvino, 1991) (in
Russian).\\
M. S. Plyushchay, A. V. Razumov, 
{\it Int. J. Mod. Phys.} {\bf A11}, n.8
(1996) 1427.

\bibitem{Storchak_11}
S. N. Storchak.  Bogolubov transformation in path integral on manifold with a group action, (IHEP Preprint 98-1, Protvino, 1998).

\bibitem{Storchak_12}
S. N. Storchak, 
{\it Reduction in path integrals on a Riemannian manifold with a group action},  
Phys. Atom. Nucl. 64, 2199 (2001).  

\bibitem{Storchak_2}
S. N. Storchak, 
{\it Dependent coordinates in path integral measure factorization}, 
J. Phys. A: Math. Gen. 
 37, 7019 (2004),
(IHEP Preprint 2000-54, Protvino, 2000), arXiv: math-ph/0311038 [math-phys].


\bibitem{Storchak_3}
S. N. Storchak, On the geometrical representation
of the path integral reduction Jacobian:
The case of dependent variables in a description of  reduced motion, {\it J. Geometry and Physics} {\bf 59}
(2009) 1155.

\bibitem{Mitter}
P. K. Mitter, Geometry of the space of gauge orbits and the Yang-Mills dynamical system.  Carg\`{e}se Lectures in {\it  Recent developments    in gauge theory}, t'Hooft.G. {\it et al} (eds.), Plenum Press, 1980.

\bibitem{Singer_2}
I. M. Singer, The geometry of the orbit space for non-Abelian gauge theories, {\it Physica Scripta} {\bf 24} (1981) 817-820.

\bibitem{Parker}
D. Groisser, T. H. Parker, The geometry of the Yang-Mills module space for definite manifolds, {\it J. Diff. Geom.} 
{\bf 29} (1989) 
499.

\bibitem{Soloviev}
Yu. P. Soloviev, Geometrical structures on a manifold of 
 interacting gauge fields, {\it Global analysis and mathematical physics} (Voronezh, 1987) 110--121 (in Russian).

\bibitem{Postnikov}
M. M. Postnikov, Lectures in geometry, Semester \Roman{4}: Differential geometry, (Nauka, Moscow, 1988) (in Russian).


\bibitem{Kunstatter} 
G. Kunstatter, {\it Class. Quant.Grav.} 
{\bf 9} (1992) 1466.

\bibitem{Falck}
N. K. Falck, A. C. Hirshfeld, {\it Ann. Phys.} {\bf 144} (1982) 34.





  
\end{thebibliography}
\end{document}